\documentclass[preprint,aps,showpacs,endfloats]{revtex4}
\usepackage{epsfig}

\newcommand{\Be}{\begin{equation}}
\newcommand{\Ee}{\end{equation}}
\newcommand{\Bea}{\begin{eqnarray}}
\newcommand{\Eea}{\end{eqnarray}}

\begin{document}

\title{Solving the Radial Dirac Equations: A Numerical Odyssey}
\author{Richard R.\ Silbar}\email
{ silbar@lanl.gov} 
\affiliation{Theoretical Division, MS-B283, 
Los Alamos National Laboratory, Los Alamos, NM 87545}
\author{T.\ Goldman}\email{ tgoldman@lanl.gov}
\affiliation{Theoretical Division, MS-B283, 
Los Alamos National Laboratory, Los Alamos, NM 87545}

\begin{flushright}
\today\\
{LA-UR-10-0063}\\
{arXiv:1001.2514}\\
\end{flushright}

\begin{abstract}
We discuss, in a pedagogical way, how to solve for relativistic wave functions from
the radial Dirac equations.  
After an brief introduction, in Section II we solve the equations for a linear Lorentz 
scalar potential, $V_s(r)$, that provides for confinement of a quark. 
The case of massless $u$ and $d$ quarks is treated first, as these are necessarily 
quite relativistic.  
We use an iterative procedure to find the eigenenergies and the upper and lower 
component wave functions for the ground state and then, later, some excited states.
Solutions for the massive quarks ($s$, $c$, and $b$) are also presented.
In Section III we solve for the case of a Coulomb potential, which is a time-like
component of a Lorentz vector potential, $V_v(r)$.
We re-derive, numerically, the (analytically well-known) relativistic hydrogen atom eigenenergies and wave functions, and later extend that to the cases of heavier 
one-electron atoms and muonic atoms.
Finally, Section IV finds solutions for a combination of the $V_s$ and $V_v$ 
potentials.
We treat two cases. 
The first is one in which $V_s$ is the linear potential used in Sec. II and 
$V_v$ is Coulombic, as in Sec. III.
The other is when both $V_s$ and $V_v$ are linearly confining, and we establish
when these potentials give a vanishing spin-orbit interaction (as has been
shown to be the case in quark models of the hadronic spectrum).

\end{abstract}

%\vspace{0.5in}

\maketitle

\section{Introduction}

Dirac formulated relativistic quantum mechanics in the late 1920's.\cite{Dirac}$\ $
Since field theory had not yet been developed, the relativistic aspect led to a number 
of confusions related to currents since only total charge is conserved. With the 
recognition of the appearance of antiparticles, and the development of field theory, 
analogous to the transition from the Canonical to the Grand Canonical ensemble in 
statistical mechanics, these issues were resolved. Nonetheless, the solution of the 
equation itself, ignoring these deeper aspects, has proven valuable even in more 
modern contexts. 

For example, the MIT bag model of quark confinement,\cite{Bag} and the theories 
that evolved from it (chiral bag, cloudy bag, etc.)\cite{AWT} depend on 
solving the Dirac equation as a wave function for a particle in an effective potential. 
Through boundary conditions and other approximations, the Dirac equation has 
even been employed in this way in the study of nuclear structure.\cite{AWT2,GMSS}$\ $ 

In particle physics, the quark model has been employed, with great success,
to describe baryon and meson states and their structure.
Usually this is done in a {\it non}-relativistic model,
\cite{CorPot} where the potentials follow the patterns expected by 
a Foldy-Wouthuysen reduction\cite{FW} of a simpler, theoretically motivated 
potential in the Dirac equation. 
In our case, however, we wanted to see how these structures and 
solutions appear {\it without} the non-relativistic reduction approximations, by solving 
the Dirac equation itself for the simpler potential with fewer adjustable parameters. 

Going back somewhat in time, very soon after Dirac's initial formulation, 
Darwin \cite{Darwin} found an analytic solution (!) for the radial wave
functions for hydrogen-like atoms in terms of confluent hypergeometric functions.
Since then, solutions of the Dirac equation have been important in atomic physics, 
even to recent times.\cite{BetheSal,AuRogers}$\ $
For example, more complicated central potentials than $Z\alpha /r$, the fourth component 
of a Lorentz four-vector, usually require finding a numerical solution.
Another example is a muonic atom, in which a muon is in an s-state about a heavy 
nucleus having a realistic charge distribution.\cite{mu_xray}

Finding such numerical solutions involves solving coupled ordinary differential 
equations for the upper and lower components of the Dirac wave function, \cite{BjD,AuRogers}
\Be
 \psi_{jlm}({\bf r}) = \left[ \begin{array}{c}
          \psi_a(r)\; {\cal{Y}}^l_{jm} \\
	   -i \psi_b(r) \; {\cal{Y}}^{l'}_{jm}
	   \end{array} \right]  = \frac{1}{r}
	   \left[ \begin{array}{c}
          i g(r)\; {\cal{Y}}^l_{jm} \\
	      -f(r) \; {\cal{Y}}^{l'}_{jm}
	   \end{array} \right]
	   \ ,
  \label{eq:psicolvec}
\Ee
where $l' = 2 j - l$ and the ${\cal{Y}}^l_{jm}$ are the spin-orbital angular momentum
wave functions,
\Be
	 {\cal{Y}}^{l'}_{jm}(\theta,\phi) = \sum_{m_s} <j m \; 
	 	|\; l \;  {\textstyle \frac{1}{2}},\; m-m_s \; m_s> 
	 	Y^l_m (\theta,\phi)\; \chi_{m_s} \ ,
\Ee
where $<j m \; |\; l \;  {\textstyle \frac{1}{2}},\; m-m_s \; m_s> $
is a Clebsch-Gordan coefficient and $\chi$ is a Pauli spinor.

We consider here the case of bound-state wave functions for a central potential with 
energy eigenvalue $E$.
For example, for hydrogen-like atoms the potential is $V_v(r) = -Z\alpha/r$.
With appropriate boundary conditions, the radial wave functions $g(r)$ and $f(r)$,
which can be taken as real functions, are solutions of the following coupled first-order 
ordinary differential equations (ODEs), \cite{BjD,BetheSal}
\Bea
	&g'(r)& + \;\frac{k}{r} g(r)\; - \;(E - V_v(r) + m)\; f(r) = 0 \ , \label{eq:Vv_geqn} \\ 
	&f'(r)& - \;\frac{k}{r} g(r)\; + \;(E - V_v(r) - m)\; g(r) = 0 \ . \label{eq:Vv_feqn}
\Eea
Here the integer $k$ is determined by the angular momentum quantum numbers according to
\Bea 
	k &=& -(l+1) {\rm \quad\quad if \ } j = l + {\textstyle \frac{1}{2}} \ , \nonumber \\
	k &=& l \quad\quad\ \ {\ \ \rm \quad\quad if \ } j = l - {\textstyle \frac{1}{2}} \ . \label{eq:kdef}
\Eea

The reason for the subscript on the potential $V_v$ in those equations is to indicate that
it is the fourth component of a Lorentz four-vector, such as the Coulomb potential.
However, {\em our} motivation is not so much in atomic physics applications as it is 
for treating mesons as $q \bar{q}$ states in a {\em relativistic} quark model.
The {\em non}-relativistic quark model often assumes the confining potential for
quarks to be the so-called ``Cornell potential,'' \cite{CorPot}
\Be 
	V(r) = -\frac{\alpha_s}{r} + \kappa^2 r \ , \label{eq:CorPot}
\Ee
where $\alpha_s = g_{s}^{\ 2}/{4\pi}$, with $g_s$ being the (running) quark-gluon coupling 
constant, and $\kappa^2$ (or $\sigma$) is the string tension.
It is the linear term in $V(r)$ that confines the quarks, 
corresponding to the effective potential in the relativistic Bag Model.\cite{Bag}

For a relativistic quark model, the two pieces of that non-relativistic potential have
different Lorentz transformation properties.
The color Coulomb potential, $\alpha_s/r$, is the fourth component of a Lorentz vector,
while the confining linear potential transforms as a Lorentz scalar,
which we will write as $V_s(r)$.
Thus, in solving the relativistic radial Dirac equations, the two terms in the non-relativistic potential of Eq.\ (\ref{eq:CorPot}) enter the coupled ODE's differently. 
The equations to be solved are 
\Bea
	&g'(r)& + \;\frac{k}{r} g(r)\; - \;(E - V_v(r) + V_s(r) + m)\; f(r) = 0 \ , 
		\label{eq:VvVs_geqn} \\ 
	&f'(r)& - \;\frac{k}{r} f(r)\; + \;(E - V_v(r) - V_s(r) - m)\; g(r) = 0 \ . 
		\label{eq:VvVs_feqn}
\Eea
Note that in these equations  the sign for $V_v$ is opposite to that of the energy
eigenvalue $E$, the fourth component of the momentum four-vector, while 
that for $V_s$ matches that for the fermion mass $m$, also a Lorentz scalar.

So, we were motivated to study solutions of Eqs.\ (\ref{eq:VvVs_geqn}) and (\ref{eq:VvVs_feqn}).  
After submitting an earlier version of this paper to the archives, 
we learned of related studies of these equations by Paris\cite{Paris} and 
by Soares de Castro and Franklin\cite{JF}, who discussed analytic solutions.
We, however, have adopted a numerical approach and,
in this paper, we discuss how to go about solving these coupled ODE's numerically.
The presentation follows our numerical journey more or less historically.
Our hope in writing this paper is that, by reading it, students might avoid some 
of the pitfalls we encountered and then overcame.

It turns out that solving for a linear scalar potential \cite{Critch,GMSS} 
is quite a bit easier (numerically) than the relativistic hydrogen atom, 
so we treat that first in Sec. II.
We then turn, in Sec. III, to the problems we had with in solving the radial equations 
for a Coulomb-like potential for hydrogen-like atoms and how we dealt with them.
This is followed, in Sec. IV, by a discussion of the mixed problem having {\it both} 
a scalar and a vector potential, as in Eqs.\ (\ref{eq:VvVs_geqn}) and (\ref{eq:VvVs_feqn}).

\section{Linear Scalar Potential}

For the case of massless quarks (an approximation we have made \cite{piExchg} 
for the $u$ and $d$ quarks),
we want to solve the equations for a confining linear 
scalar potential, \cite{Critch,GMSS}
\Be
   V_s(r) = \kappa^2 (r - r_0) \; , \label{eq:GMSSpotnl} 
\Ee
where (with $\hbar = c = 1$), $\kappa$ has dimensions of fm$^{-1}$. 
The negative offset, $-\kappa^2 r_0$, effectively gives these quarks a constituent mass.\cite{GMSS}$\ $
It also provides a rough representation of the effect of how the short-distance color 
Coulomb interaction between quarks leads to quark confinement, albeit without the
correct Lorentz representation properties.

The case of massless quarks exhibits the role of relativity in a maximal way, meaning
that the lower component $f(x)$ is comparable in size to the upper component $g(x)$.
We will, later, choose an appropriate mass $m$ for the massive quark flavors, $s$, $c$, and $b$,
and solve for those wave functions.
In these cases, as the quark mass $m$ increases, $f(x)$ becomes smaller relative to $g(x)$,
reflecting the transition to a non-relativistic limit where $f(x)$ vanishes.
Also, as will be seen, with increasing mass $m$, the heavier quark radial wave functions 
are progressively narrower than in the massless case.
Thus we will, in Sec.\ \ref{VsAndVv}, also include the effects of a color
Coulomb-like vector potential $V_v$ coming from one-gluon exchange.
This should prevent a too-rapid fall-off of the 
quark-quark interaction energies as the quark masses increase \cite{AprilTalk}, 
but that is a matter for a separate paper.

It simplifies the equations if we convert the ODEs to dimensionless form, 
dividing through by $\kappa$ and defining a dimensionless distance $x = \kappa r$:
\Bea
	&g'(x)& + \;\frac{k}{x} g(x)\; - \;(\tilde{E} + V_s(x) + \tilde{m})\; f(x) = 0 \ , 
		\label{eq:Vs_geqnx} \\ 
	&f'(x)& - \;\frac{k}{x} f(x)\; + \;(\tilde{E} - V_s(x) - \tilde{m})\; g(x) = 0 \ , 
		\label{eq:Vs_feqnx}
\Eea
where $\tilde{E} = E/\kappa$, $V_s(x) = x-x_0$, and $\tilde{m} = m/\kappa$ 
are now also dimensionless.

Numerical integrations of these equations can be done by one or another of the 
Runge-Kutta methods.\cite{RK}$\ $
To integrate these two first-order ODEs one needs to specify two boundary conditions (BCs)
at the place where one begins the integration.
However, as the radial wave functions eventually need to satisfy
a normalization condition,
\Be 
	\int_{0}^{\infty} \; x^2 dx \;  [\psi_a^2(x) + \psi_b^2(x)] = 
	\int_{0}^{\infty} \; dx \;  [g^2(x) + f^2(x)] = 1 \ , \label{eq:normzn}
\Ee
this will determine one of the BCs, for example, the scale of the wave functions at 
asymptotically large distances.

\subsection{Shooting Outwards}

In our first attempt to solve these equations we thought to integrate them outward 
from the origin  
with the hope of adjusting the energy eigenvalue $\tilde{E}$ to assure
that both $g(x)$ and $f(x)$ fall off to zero asymptotically.  
The first problem in doing this is the singularity in the equations at $x=0$.
This singularity, however, is easily avoided by starting instead at a small value 
of $x = \epsilon$ away from the origin.

For starting values (BCs) we can take advantage of the expected 
power-law behaviors of $\psi_a(x)$ and $\psi_b(x)$ near the origin for their 
known orbital angular momenta $l$ and $l'$.
For example, if $j = l + \frac{1}{2}$, for which $k = -(l+1)$ and the lower 
component's $l' = l+1$, one can choose
\Be 
	g(\epsilon) = a_0 \;\epsilon^{l+1} \ , \quad
	f(\epsilon) = b_0 \;\epsilon^{l+2} \ , \label{eq:BC34_l+1/2}
\Ee
as motivated by the non-relativistic limit, where 
$\psi_a \rightarrow \Psi_{\rm NR} \sim r^l$ at small $r$.
Substituting this $f(\epsilon)$ into the second ODE allows us to determine $b_0$:
\Be 
	f(\epsilon) = -a_0 \; (\tilde{E} - V_s(\epsilon) - \tilde{m})\; 
		\epsilon^{l+2}/(2 l + 3) \ . \label{eq:fBC4}
\Ee
The parameter $a_0$ at this point is arbitrary for now and will be fixed later 
by the normalization condition, Eq.\ (\ref{eq:normzn}).

The problem with simply shooting outwards is what to choose for the energy $E$.
The ODE solver for almost all initial guesses for $E$ will have the 
well-known problem that the calculated
$g(r)$ and $f(r)$ very soon blow up exponentially, either positively or negatively. 
It might be possible to iteratively refine the guess for $E$ to eventually find 
decaying solution, but we soon decided to try a different method.

\subsection{Shooting Inwards}

A better way of assuring decaying solutions is to start from an asymptotic distance,
where that behavior is built-in, and integrate the ODEs inward toward the origin.  
At large distances the ODEs simplify, for this potential, to
\Bea
	&&g'(x) -  x f(x) = 0 \ , \nonumber \\
	&&f'(x) -  x g(x) = 0 \ ,
\Eea
which have solutions at large $x = x_{\rm max}$,
\Bea
	&&g(x_{\rm max}) =  a_1 e^{-x_{\rm max}^2/2} \ , \nonumber \\
	&&f(x_{\rm max}) = -a_1 e^{-x_{\rm max}^2/2} \ . \label{eq:BC12}
\Eea
Note that this asymptotic behavior is the same as that of the non-relativistic 
simple harmonic oscillator wave functions.\cite{Critch}$\ $
These forms will be used as starting values (BCs) for the inward integration of the 
full ODEs from $x_{\rm max}$.
Again, $a_1$ can be taken as arbitrary for now, to be later fixed by the normalization 
condition, but we need to make an initial guess for the energy eigenvalue $\tilde{E}$.
The integration will go from $x_{\rm max}$ back to $x = \epsilon$, again to avoid
the singularity at $x=0$.

The problem with this shooting inwards method is much the same as that for
shooting outwards.
Unless one somehow is able to guess the exact value of $\tilde{E}$,
the solutions will blow up exponentially as one nears the
origin, and usually well before that.

\subsection{Shooting In and Out and Matching}

Thus, after experiencing these well-known problems with ODEs with eigenvalues,
we concluded that we had to combine the two methods, invoking a well-known
numerical method, ``shoot and match.''
In this case we shoot both outwards from $x = \epsilon$ to a point in the middle, 
$x_{\rm match}$, and inwards from $x_{\rm max}$ back to $x_{\rm match}$.
The two values for $g_{\rm out}(x_{\rm match})$ and $g_{\rm in}(x_{\rm match})$
will generally differ for a given choice of $\tilde{E}$ and another parameter, 
which we choose to be $a_0$.
That is likewise the case for $f_{\rm out}(x_{\rm match})$ and $f_{\rm in}(x_{\rm match})$.
What we therefore need is to find a way to {\it iteratively vary} $\tilde{E}$ and $a_0$
to reduce the two gaps, $\Delta g$ and $\Delta f$, to zero.
[The asymptotic parameter $a_1$ is chosen to be fixed, i.e., not to be varied, 
but of a size that makes the initial determination of the gaps at $x_{\rm match}$ 
reasonably small; it will in the end be determined by the normalization condition,
Eq.\ (\ref{eq:normzn}).]

We need to define the gaps $\Delta g$, etc., at the match point.  
Initially, we simply chose them in terms of their actual differences,
\Be 
	\Delta g(\tilde{E},a_0) = g_{\rm out}(x_{\rm match}) - g_{\rm in}(x_{\rm match})
	\ , \label{eq:actualgaps}
\Ee
and likewise for $\Delta f$, $\Delta g'$, and $\Delta f'$.
Indeed, this often works for the problem we originally set out to do, namely, to calculate 
wave functions for the (nearly) massless $u$ and $d$ quarks.
However, as the quark mass $\tilde{m}$ increases, the system becomes more and more 
non-relativistic.
This means that the lower component wave function, $f(x)$, becomes relatively small
compared with $g(x)$.  
Thus a better definition for the gaps would be to make them relative to their average
value, e.g.,
\Be 
	\Delta g(\tilde{E},a_0) = 
		\frac{2\;[\;g_{\rm out}(x_{\rm match}) - g_{\rm in}(x_{\rm match})\;]}
			{g_{\rm out}(x_{\rm match}) + g_{\rm in}(x_{\rm match})}
		\ , \label{eq:scaledgaps}
\Ee
and likewise for $\Delta f(\tilde{E},a_0)$.

\subsection{Zeroing the Gaps}

The two gaps can be driven to zero by using a generalized Newton-Raphson method.
First, recall how one derives the Newton-Raphson method for finding the zero
of a function $y(x)$.  
Make an initial guess for the solution, call it $x_1$.
To find a better guess, $x_2$, expand in a Taylor's expansion,
\Be 
	y(x_2) = y(x_1) + y'(x_1) (x_2 - x_1) + \cdots \ . \label{eq:Taylor}
\Ee
As we want the left-hand side to vanish, the next (and better) guess for the root is
\Be 
	x_2 = x_1 - y(x_1)/y'(x_1) \ .  \label{eq:Newton}
\Ee
This procedure can be repeated until $y(x_n)$ is small enough to be considered zero.

The generalization of Eqs.\ (\ref{eq:Taylor}) and (\ref{eq:Newton}) for our problem is 
to solve a two-by-two linear system for new values of the parameters, 
$\tilde{E}_{\rm new}$ and $a_{0,{\rm new}}$,
from these equations:
\Bea
	0 &=& \Delta g +  \left(\frac{\partial\;\Delta g}{\partial\tilde{E}}\right) 
	(\tilde{E}_{\rm new} - \tilde{E}_{\rm old}) +
	\left(\frac{\partial\;\Delta g}{\partial a_0}\right) (a_{0,{\rm new}} - a_{0,{\rm old}}) 
	\nonumber \\
	0 &=& \Delta f +  \left(\frac{\partial\;\Delta f}{\partial\tilde{E}}\right) 
	(\tilde{E}_{\rm new} - \tilde{E}_{\rm old}) +
	\left(\frac{\partial\;\Delta f}{\partial a_0}\right) (a_{0,{\rm new}} - a_{0,{\rm old}})
	\ . \label{eq;delgf}
\Eea

We will calculate the partial derivatives needed above numerically.
The procedure to find better values of the two parameters $\tilde{E}_{\rm new}$ and 
$a_{0,{\rm new}}$ is to be iterated until the gaps are sufficiently small.

To solve this linear system in Eq.\ (\ref{eq;delgf}), it is convenient to define 
a matrix
\Be 
	M = \left[ \begin{array}{cccc}
	\frac{\partial\;\Delta g}{\partial\tilde{E}} & 	\frac{\partial\;\Delta g}{\partial a_0}\\
	\frac{\partial\;\Delta f}{\partial\tilde{E}} & 	\frac{\partial\;\Delta f}{\partial a_0}\\	
	\end{array} \right] \ \label{eq:2x2partials}
\Ee
and a column vector containing the gaps and partials obtained using the old parameters 
$\tilde{E}_{\rm old}$ and $a_{0,{\rm old}}$,
\Be
	C_{\rm old} = \left[ \begin{array}{c}
	\left(\frac{\partial\;\Delta g}{\partial\tilde{E}}\right) \tilde{E}_{\rm old} 
		+ \left(\frac{\partial\;\Delta g}{\partial a_0}\right) a_{\rm old} - \Delta g_{\rm old} \\
	\left(\frac{\partial\;\Delta f}{\partial\tilde{E}}\right) \tilde{E}_{\rm old}
		+ \left(\frac{\partial\;\Delta f}{\partial a_0}\right) a_{\rm old} - \Delta f_{\rm old} 
	\end{array} \right] \ . \label{eq:2x2oldC}
\Ee
Multiplying $C_{\rm old}$ by the inverse matrix $M^{-1}$ then provides us with a 
column vector containing the improved parameters for the next iteration:
\Be 
	P_{\rm new} = M^{-1} \; C_{\rm old} 
		= [\tilde{E}_{\rm new}, a_{0,{\rm new}}]^T 
		\ . \label{eq:2x2Pnew}
\Ee

It turns out that this iterative procedure works very well, 
with the gaps in the {\it slopes} of $g(x)$ and $f(x)$ at the match point 
{\it also} going to zero.

At this point we should remark that others might have done this calculation
but by zeroing the gaps in the logarithmic derivatives, such as$(dg(x)/dx)/g(x)$.
This has the advantage of removing the scale dependence of the functions $g$ and $f$,
but our choice of defining the gaps as relative, as in
Eq.\ (\ref{eq:scaledgaps}), is essentially equivalent, as 
it also involves (semi-)local differences divided by the (semi-)local value.

\subsection{The $1s$ Radial Wave Functions and Some Programming Details \label{sec:Vs1s}}

We implemented the iterative numerical process discussed above
by developing Mathematica notebooks, \cite{MMa} but any reasonable programming language 
could be used instead.
An example notebook is available from our group's web site.\cite{RRSwebpage}$\ $ 
(We especially invite our Canadian colleagues to convert this to a Maple program!)
Here, 
as we describe the calculation of the ground state radial wave functions for
the GMSS potential of Eq.\ (\ref{eq:GMSSpotnl}), we will comment on a number of the programming issues that we encountered.

The first thing to be done is to set the value of $k$ in the equations to be solved.
It is an integer and depends upon the angular momentum quantum numbers, Eq.\ (\ref{eq:kdef}).
For the $1s$, $j=1/2$ ground state, $l = 0$, $l' = 1$, and thus $k = -1$.

As noted earlier, we work in units where $\hbar = c = 1$ \cite{hbarc} so that the 
dimensions of Eqs.\ (\ref{eq:VvVs_geqn}) and (\ref{eq:VvVs_feqn}) are fm$^{-1}$.
For the cases discussed here and below, we have fixed the constants in the GMSS potential 
at $\kappa^2 = 0.9 {\rm \ GeV/fm}$ (i.e., $\kappa = 2.14 {\rm\ fm^{-1}}$),
and $r_0 = 0.705$ fm.
Making the equations dimensionless, as mentioned after Eq.\ (\ref{eq:Vs_feqnx}),
this linear scalar potential simplifies to $\tilde{V}_s(x) = x - x_0$, 
where $x_0 = \kappa r_0 = 1.506$.

For dimensionless distances we chose $\epsilon = 10^{-6}$, $x_{\rm match} = 1.0$, and 
$x_{\rm max} = 6.0$.
We also define a small increment $\delta = 0.0001$, which will be used 
when we calculate the partial derivatives in Eq.\ (\ref{eq:2x2partials}) numerically
and for testing when the gaps to be zeroed are small enough.

The two integrations, inwards and outwards, were done using Mathematica's {\tt NDSolve} 
function, but one could use any standard Runge-Kutta procedure.\cite{RK}$\ $
To proceed, we need the four boundary conditions (BCs) from Eq.\ (\ref{eq:BC34_l+1/2}) 
and Eq.\ (\ref{eq:BC12}).
We have defined a subroutine, {\tt  shootinandout}, to do this.  
This subroutine is, of course, a function of the parameters $\tilde{E}$ and $ a_0$.
It saves the results of the integrations as {\tt insoln} and {\tt outsoln}, respectively.

In this case, we already had a good idea \cite{GMSS} of what the energy eigenvalue $\tilde{E}$ 
is for the $1s$ ground state.
After some fiddling, we found an initial choice of parameters
\Be 
	\tilde{E} = 0.82, \quad a_0 = 0.2, \quad a_1 = 1000.0 \label{eq:params1sVs}
\Ee
for which the resulting outwards and inwards integrations yielded the curves for 
$g(x)$ and $f(x)$ shown in Fig.\ \ref{fig:initialVs1s}.
(As mentioned above, the asymptotic normalization $a_1$ is at this point arbitrary, to be fixed later by the normalization condition.)

It is useful at this point to define a function, {\tt calcmatchgaps}, that uses {\tt insoln} 
and {\tt outsoln} to calculate and print out the values of $g(x)$ and $f(x)$ 
at the match point $x_{\rm match}$, along with their slopes and their gaps scaled as in
Eq.\ (\ref{eq:scaledgaps}).
These numbers provide guidance as to how to proceed when developing the Mathematica notebook.

Next, we need the four partial derivatives in the two-by-two matrix $M$ of 
Eq.\ (\ref{eq:2x2partials}). 
In pseudocode, the subroutine for calculating ${\partial\;\Delta g/\partial\tilde{E}}$ is

{\samepage
{\baselineskip=12 pt
\begin{verbatim}
   subroutine dggapbydE(tildeE,a0) 
      delE = delta*tildeE 
      outsoln1 = outward integration with parameters (tildeE + delE,a0) 
                 from epsilon to xmatch
      g_out1 = g(x) at x_match from outsoln1
      insoln1 = inward integration with parameters (tildeE + delE,a0) 
                from xmax to xmatch
      g_in1 = g(x) at x_match from insoln1
      ggap1 = 2.0*(g_out1 - g_in1)/(g_out1 + g_in1)
      dggapdE = (ggap1 - ggap)/delE
      return dggapdE
      end subroutine
\end{verbatim}
}}
\vspace{-12 pt}\noindent 
Similar subroutines are also implemented for the three other partial derivatives.
We also found it useful, once these subroutines were in place, to define another subroutine, 
{\tt gapsandpartials}, which calls {\tt shootinandout} for the present values of the  
parameters $\tilde{E}$ and $a_0$, followed by calls of {\tt calcmatchgaps} and the 
four subroutines for the partial derivatives.

At this point we are ready to solve the two-by-two linear system for improved values 
of the parameters $\tilde{E}$ and $a_0$.
Manipulating arrays is a bit tricky in a programming language such as Fortran or C,
but in Mathematica one can simply define the matrix $M$ in Eq.\ ({\ref{eq:2x2partials}) as a list of lists and the column vector $C_{\rm old}$ is a simple list. 
Mathematica also provides a built-in function to invert the matrix $M$, 
so the evaluation of the right-hand-side of Eq.\ (\ref{eq:2x2Pnew}) is easy.
This procedure is encapsulated in a subroutine we called {\tt solvelinsys()}. 

Before going on to the program for the iterative loop we need to decide when
the gaps are small enough to stop the iteration.  
Thus we defined a test function which returns a boolean
{\tt True} if {\em any} of the {\it four} gaps, $\Delta g$ and $\Delta f$ as well as their slopes 
$\Delta g'$, and $\Delta f'$, is greater than $\delta$.
Again, in pseudocode,

{\samepage
{\baselineskip=12 pt
\begin{verbatim}
   subroutine testgaps() 
      boolean p1 = Abs[ggap] > delta
      boolean p2 = Abs[fgap] > delta
      boolean p3 = Abs[gpgap] > delta
      boolean p4 = Abs[fpgap] > delta
      boolean q = p1 OR p2 OR p3 OR p4 
      return q
      end subroutine 
\end{verbatim}
}
}

\vspace{-12 pt}\noindent 
%{\bf (Check this!)}
The iteration will stop when {\tt q = False}.

The coding for running the iteration from an initial choice of parameters 
$\tilde{E}$ and $a_0$ is as follows:

{\samepage
{\baselineskip=12 pt
\begin{verbatim}
   {newtildeE,newa0} = {start_tildeE, start_a0}
   print starting parameters {newtildeE, newa0, a1}
   iter = 0
   q = True
   do loop
      iter = iter+1
      {oldtildeE,olda0} = {newtildeE,newa0}
      shootinandout(oldtildeE,olda0)
      gapsandpartials(oldtildeE,olda0)
      solvelinsys()   // returns improved values of newtildeE and newa0
      print iter and gaps
      print newtildeE and newa0
      if q
         continue to next iteration, i.e., go to the top of the do loop
      else 
         stop and break out of the do loop
      end if
      end do loop 
\end{verbatim}
}
}

\vspace{-12 pt} 

For the starting parameters used above, $\tilde{E} = 0.82$ and $a_0 = 0.2$,
we find that our program closes the gaps  in four iterations.
The final parameter values are $\tilde{E} = 0.727102$ and $a_0 = 0.194709$.
That $\tilde{E}$ correponds, in more conventional units, to an energy of $E = 0.306$ GeV.
Figure 2 shows plots of the final, converged $g(x)$ and $f(x)$ after normalization.
Note the relatively large size of the lower component $f(x)$, showing the importance of 
relativity in this case of massless quarks.

Figure 3 shows the corresponding plots for $\psi_a(x)$ and $-\psi_b(x)$.
As one ought expect for a ground state wave function,
the upper component $\psi_a(x)$ has no nodes and the $p$-wave lower component 
$\psi_b(x)$ has one (at the origin).

We have plotted the $\psi$'s in Fig.\ 3 this way to compare with GMSS's Fig.\ 2.
The $\psi$'s here are for fitting the non-strange $q\bar{q}$ mesons and extend 
out further than those of GMSS.
The reason for that is that GMSS calculated the $u$ and $d$ quark wave functions to fit the non-strange baryon spectrum, which is why they used a value of $r_0 = 0.57$.
Physically, the effective origin of the confining potential for mesons is 
not as strongly localized as in the baryon case, where one more quark damps 
fluctuations additionally.

\subsection{Some Excited States \label{VsExciteds}}

The procedure outlined in the previous sub-section can be applied to calculate the energy 
eigenvalues and wave functions for excited states.
One expects that the first excited state is the $2s\;\frac{1}{2}$ state.
This state also has $l = 0$, $l' = 1$ and $k = -1$, so the BCs at $x = \epsilon$ are the same 
as for the $1s$ case.
The upper component wave function $g(x)$ should now have a new node, i.e., cross the $x$-axis
somewhere between the origin and infinity.
Thus, if it is desirable to have $g(x)$ start off from the origin going positive,
one should choose the asymptotic normalization parameter $a_1$ to be negative.

In view of the asymptotic behavior here being similar to that of a simple 
harmonic oscillator, we might expect that the energy eigenvalue of this state is 
roughly twice that of the ground state.
Choosing the parameters needed for the initial integrations to be
\Be 
	\tilde{E} = 2.1, \quad a_0 = 0.3, \quad a_1 = -4000.0 \ , \label{eq:params2sVs}
\Ee
the iterative loop again converges in four iterations. 
The final parameters are $\tilde{E} = 1.91897$ (i.e., 0.809 GeV) and $a_0 = 0.378981$.
Plots of $\psi_a(x)$ and $-\psi_b(x)$ are displayed in Fig.\ 4.
For the $2s\;1/2$ state, $\psi_a(x)$ has one node and $\psi_b(x)$ has two.

Why are we counting nodes, anyway?  Because a state with more nodes has more energy.
Qualitatively, more nodes means more curvature, and that means a bigger contribution from 
$\nabla^2$, which in turn means a bigger kinetic energy.  

Besides the $2s\;\frac{1}{2}$ state, there are also two nearby $p$-wave excited states.  
The $2p\;\frac{3}{2}$ state also has $j = l + \frac{1}{2}$, but now with $l = 1$, $l' = 2$, and thus $k = -2$.
The BCs near the origin are given by Eqs.\ (\ref{eq:BC34_l+1/2}) and (\ref{eq:fBC4}), 
but otherwise the coding
is very similar to (basically can be copied from) the $1s\;\frac{1}{2}$ and 
$2s\;\frac{1}{2}$ cases.
Starting this time with
\Be 
	\tilde{E} = 2.1, \quad a_0 = 3.0, \quad a_1 = -4000.0 \ , \label{eq:params2p3by2Vs}
\Ee
the gaps close in four iterations, with final parameters $\tilde{E} = 2.23003$ 
(i.e., 0.940 GeV) and $a_0 = 1.57784$.
The $2p\;\frac{3}{2}$ wave functions $\psi_a(x)$ and $-\psi_b(x)$ are shown in Fig.\ 5.
In contrast to the $2s$ case, here {\it both} $\psi_a(x)$ and $\psi_b(x)$ have two nodes,
reflecting a higher energy eigenvalue.

The other $p$-wave state, $2p\;\frac{1}{2}$, is qualitatively different.
This is the first case for which $j = l - 1/2$, and $k$ is now a positive integer.
In a sense, this switches the roles of $g(x)$ and $f(x)$.
The boundary conditions for the outward integration when $j = l - \frac{1}{2}$ are 
different from those of Eq.\ (\ref{eq:BC34_l+1/2}):
\Be 
	f(\epsilon) = a_0 \epsilon^l \ , \quad
	g(\epsilon) = b_0 \epsilon^{l+1} \ . \label{eq:BC34_l-1/2}
\Ee
Substituting $g(\epsilon)$ into the first ODE allows us to determine this $b_0$, so
\Be 
	g(\epsilon) = a_0 \; (\tilde{E} + V_s(\epsilon) + \tilde{m}) 
		\epsilon^{l+1}/(2 l + 1) \ . \label{eq:gBE4}
\Ee
That is, if $a_0 >0$, {\it both} $g(x)$ and $f(x)$ start out from the origin with 
positive slopes.

Starting for this $2p\;\frac{1}{2}$ case with initial parameters
\Be 
	\tilde{E} = 2.1, \quad a_0 = 3.0, \quad a_1 = -4000.0 \ , \label{eq:params2p1by2Vs}
\Ee
the iterative procedure converges on the fifth iteration, with final parameter values
$\tilde{E} = 2.37846$ (i.e., 1.002 GeV) and $a_0 = 0.171483$.
The $\psi$'s are plotted in Fig.\ 6.
Here also both component wave functions have two nodes. 

%{\bf DISTURBING: the 2p states should be below the 2s?????}

We note in passing that the higher spin state of this pair has the lower energy, 
contrary to the well-known case for the Coulomb potential. This is a common 
feature for nuclear states (even ground states) and reflects the presence of an 
effective scalar potential as found in many nuclear potential models.\cite{Walecka}

The energy difference between the $2p\;\frac{3}{2}$ and $2p\;\frac{1}{2}$ states, 
62 MeV, is due to a spin-orbit interaction.
However, there is evidence in the meson spectrum that the spin-orbit interaction
is suppressed.\cite{Isgur} 
Page, Goldman, and Ginocchio (PGG) \cite{PGG} claim this reflects a 
cancellation between a scalar potential $V_s$ and a vector potential, $V_v$, having
the same linear slope at large distances.  
We will return to this point in Sec.\ \ref{VsAndVv}.

\subsection{Solutions for Massive Quarks}

The cases for the linear scalar potential $V_s(r)$ of Eq.\ (\ref{eq:GMSSpotnl})
when the quarks are massive are computed straightforwardly using the program
described above.
For example, using masses appropriate \cite{Qqbar} for the $Q\bar{q}$ mesonic states,
where $q$ stands for a massless non-strange quark ($u$ and $d$) and $Q$ for the 
massive strange ($s$), charmed ($c$), and bottom ($b$) quarks, 
we find the following eigenenergies:

\Bea
	m_u = m_d = 0.0  \  {\rm GeV}\ , \quad & & E_u = E_d = 0.306 \  {\rm GeV} \ , \label{eq:m_ud} \\
	m_s = 0.3445  \  {\rm GeV} \ , \quad & & E_s = 0.486 \  {\rm GeV} \ , \label{eq:m_s}\\
	m_c = 1.803  \  {\rm GeV} \ , \quad & & E_c = 1.667 \  {\rm GeV} \ , \label{eq:m_c} \\
	m_b = 5.298  \  {\rm GeV} \ , \quad & & E_b = 5.007 \  {\rm GeV} \label{eq:m_b} \ ,
\Eea
by requiring a match to the appropriately weighted average of the lowest 
pseudoscalar and vector states.\cite{PDG}
Note that for the heaviest quarks the states are conventionally bound,
$E_Q < m_Q$, and confinement need not be invoked to understand the stability
of the state, in contrast to the situation for the light quarks.

Figure \ref{fig:upperWFs} compares the results for the $u$, $s$, $c$, and $b$ quarks 
for the upper components, $\psi_a(x)$, and Fig.\ \ref{fig:lowerWFs} does the same for
the lower components, $-\psi_b(x)$.
Note that as the mass $m_q$ increases, the $Q\bar{q}$ system becomes 
more and more non-relativistic,
i.e., the lower component wave function $\psi_b$ gets smaller, relative to the
upper component $\psi_a$.
Also, as $m_q$ gets larger, the wave functions are more and more concentrated
near the origin.

%\pagebreak
\section{Vector Potential -- Hydrogen-like Atoms \label{sec:hatom}}

In contrast with the case of massless quarks, an electron bound in a Coulomb potential,
as in the hydrogen atom, is at the opposite extreme, i.e., is
to a very good approximation a completely non-relativistic system.  
The binding energy of the ground state, Ry = 13.6 eV, is very small compared to the 
mass of the electron, $m_e =  0.511$ MeV.
(In this section we will use MeV instead of GeV.)
The Schr\"odinger equation for this problem predicts its energy levels basically correctly, 
missing only the fine-structure splitting, a spin-orbit effect about $10^{-4}$ times 
smaller than the binding of the $n = 2$ levels.\cite{Eisberg}.

As noted in the Introduction, the analytic solutions of the Dirac equation for the 
hydrogen atom were found long ago, \cite{Darwin} and these solutions do give the 
fine-structure splitting as a relativistic effect.
Nonetheless, it is an interesting numerical exercise to see if the program for the 
linear scalar potential outlined above (or something like it) can be applied to the 
hydrogen atom case, and with enough accuracy.
We want to solve Eqs.\ (\ref{eq:Vv_geqn}) and (\ref{eq:Vv_feqn}) with 
$V_v(r) = -Z \alpha /r$.
We will set $Z=1$, as these solutions are somewhat more delicate.
Heavier, hydrogen-like atoms with $Z>1$ can be done in a similar manner.

Converting to dimensionless equations by dividing through by $m_e$ (instead of $\kappa$ as before),
\Bea
	&g'(x)& + \;\frac{k}{x} g(x)\; - \;(1 + \tilde{E} + Z\alpha /x)\; f(x) = 0 \ , \label{eq:Vv_geqnx} \\ 
	&f'(x)& - \;\frac{k}{x} f(x)\; - \;(1 - \tilde{E} - Z\alpha /x)\; g(x) = 0 \ , \label{eq:Vv_feqnx}
\Eea
where now $x = m_e r$ and $\tilde{E} = E/m_e$.
The energy $E$ is better written as $E = m_e - B$, where $B$ is the binding
energy of the level of interest (which has principal quantum number $n$ and orbital angular momentum quantum number $l$).
Note that, because of the smallness of the binding energies, $E$ is only slightly less than $m_e$.
Thus $\tilde{E}$ is less than (but very close to) 1, which is why
we factored out a minus sign 
from the third term in Eq.\ (\ref{eq:Vv_feqnx}), relative to that in Eq.\ (\ref{eq:Vv_feqn}).
We will use the binding energy $\tilde{B} = B/m_e$ instead of $\tilde{E}$ as one of 
the two parameters to be determined in the matching of the inwards and outwards integrations.

\subsection{Boundary Conditions for Inward Integration \label{inBCs}}

For asymptotically large $x$ the ODEs reduce to
\Bea
	g'(x) &=& (1 + \tilde{E}) \; f(x)  \ , \nonumber \\ 
	f'(x) &=& (1 - \tilde{E}) \; g(x) \ . \label{eq:asympteqsH}
\Eea
The solutions of these asymptotic equations at $x = x_{\rm max}$ are
\Bea
	g(x_{\rm max}) &=& a_1 \; e^{-\mu x_{\rm max}}  \ , \nonumber \\ 
	f(x_{\rm max}) &=& -a_1 \left( \frac{1-\tilde{E}}{1+\tilde{E}}\right)^{1/2} 
		e^{-\mu x_{\rm max}}  \ ,  \label{eq:asymptsolnsH}
\Eea
which will be used as the BCs (starting values) for the inwards integration with $a_1$
as another parameter to be determined later by the normalization condition.
The coefficient in the exponential decay is small: 
\Be 
	\mu = \sqrt{1 - \tilde{E}^2} = \sqrt{2\tilde{B} - \tilde{B}^2}
	\approx\sqrt{2\tilde{B}} \; \leq \; 0.00730 \approx \alpha
	\ . \label{eq:mudef}
\Ee
The Coulomb wave functions are quite long-ranged, and are all the more so the
smaller the binding energy.
Equation (\ref{eq:asymptsolnsH}) also shows that, asymptotically, $f(x)$ is very much
smaller than $g(x)$.
The smallness of the lower component is an indication of how non-relativistic 
the hydrogen atom is.

\subsection{Boundary Conditions for Outward Integration}

As before, to avoid the singularity at $x=0$, we integrate outwards from 
$x=\epsilon$, where $\epsilon$ takes an appropriately small value.
In this case, one might even consider taking $\epsilon$ as the radius of the
nucleus providing the Coulomb potential (but see Sec. \ref{muonic} for a
better approach).
For the boundary conditions for integrating outwards from $x = \epsilon$ when
$j = l + \frac{1}{2}$, i.e., $l' = l+1$ and $k = -(l+1)$, we can again assume
\Be 
	g(\epsilon) = a_0 \;\epsilon^{l+1} \ , \quad f(\epsilon) = b_0 \;\epsilon^{l+2} \ . 
\Ee
Substituting these in the second ODE, we can solve for $b_0$, finding
\Be 
	f(\epsilon) = - a_0 \frac{Z\alpha}{2 l + 3}\;\epsilon^{l+1} + {\cal O}(\epsilon^{l+2}) 
	\ . \label{eq:Vvlower_f}
\Ee
Noticeably different is the lowering of the power behavior of $f(x)$ due to the
$1/x$ behavior of the potential.
Note that, because of the factor of $\alpha \approx 1/137$, the magnitude of $f(x)$
is much smaller than that of $g(x)$, also here near the origin.

For the $j = l - \frac{1}{2}$ case, i.e., $l' = l-1$ and $k = l$, we let $f(x)$ determine the
nature of the boundary conditions [as in Eq.\ (\ref{eq:BC34_l-1/2})], i.e.,
\Be 
	f(\epsilon) = a_0 \;\epsilon^l \ , \quad\quad g(\epsilon) = b_0 \;\epsilon^{l+1} \ . 
\Ee
Substituting these in the first ODE, we again solve for $b_0$ and find
\Be 
	g(\epsilon) = a_0 \frac{Z\alpha}{2 l + 1} \epsilon^l + {\cal O}(\epsilon^{l+1}) \ .
\Ee
Again, for this case, $g(x)$ and $f(x)$ start off with the same slope, as in 
Eq.\ (\ref{eq:gBE4}).
And again, there is a lower power behavior of $g(x)$ due to the $1/x$ in the potential, 
as in Eq.\ (\ref{eq:Vvlower_f}).

\subsection{The $1s$ Ground State \label{sec:H1sgs}}

The natural unit of length for the hydrogen atom problem is the Bohr radius, 
$a_B = \hbar ^2/m_e \alpha = 0.529$ \AA $\ = 0.529 \times 10^5$ fm.
Thus a natural scale for $x$ is in units of the dimensionless Bohr radius,
$x_B = a_B m_e/\hbar c$ which is $1/\alpha$ in $\hbar = c = 1$ units.

In view of the slow exponential decay in Eq.\ (\ref{eq:asymptsolnsH}) (i.e., 
the smallness of $\mu$), for our calculations we chose $x_{\rm max} = 7 x_B$, along with 
$x_{\rm match} = 0.5 x_B$.
Fixing $a_1 = 10$ (recall, it is arbitrary before normalization) 
and the starting values of the parameters to be tuned,
$B = 12\times10^{-6}$ MeV (less than the expected 13.6 eV) and $a_0 = 0.015$, 
we found the $2 \times 2$ matrix of partial derivatives to be
\Be 
	M = \left[ \begin{array}{rr}
	  -108877 & 107.348 \\
	  -320810 & 107.494  \\	
	\end{array} \right] \ . \label{eq:Hpartials}
\Ee
This shows a big sensitivity of the hydrogen atom wave functions to the 
choice of binding energy.
Nonetheless, the iterative process of refining $\tilde{B}$ and $a_0$ proceeds nicely to
a solution with $B = 13.6059$ eV and $a_0 = 0.0104306$.
The wave functions $g(x)$ and $f(x)$ are displayed as the solid curves in 
Figs.\ \ref{fig:Huppers} and \ref{fig:Hlowers}.
As expected, $f(x)$ is much smaller than $g(x)$.
Their shapes differ from those shown for the scalar linear potential shown 
in Fig.\ \ref{fig:finalVs1s}.

Figures \ref{fig:Huppers} and \ref{fig:Hlowers} also display, as dashed curves, 
the corresponding $\psi_a(x)$ and $\psi_b(x)$ radial wave functions.
These are, in fact, both purely decaying exponentials proportional to $e^{-\mu x}$.
This is not surprising for the upper component, $\psi_a(x)$, since that
is the just what the Schr\"odinger equation predicts for the hydrogen atom ground state.
It may not be so obvious, however, that the lower component, $\psi_b(x)$, has the
same form.
That it has no node at the origin (in contrast to Fig.\ \ref{fig:finalVs1sPsis}) is a
consequence of the $Z\alpha/x$ potential changing the presumed $x^{l+2}$-dependence
of $f(x)$ near the origin to an $x^{l+1}$-dependence.

\subsection{Some Excited States}

The $n=2$ excited states of hydrogen are done in a very similar manner to the 
$1s$ ground state calculation.  
The major difference is that, from the Balmer formula \cite{Eisberg}, we expect the binding 
energy $B$ for these states to be about $1/n^2 = 1/4$ times the ground state binding energy.
Also, based on Eq.\ (\ref{eq:mudef}),
we expect the wave functions will extend outward about twice as far, 
so, for our calculations we chose $x_{\rm max} = 16 \; a_B$ instead of $7 \; a_B$.

For the $2s\;\frac{1}{2}$ state, starting with initial parameters $B = 3.2$ eV (presumably low), 
$a_0 = 0.0007$, and (fixed) $a_1 = -10.0$,
the program converges in four iterations to give final $B = 3.40085$ eV and 
$a_0 = 0.00071905$.
The $2s$ wave functions are plotted in Figs.\ \ref{fig:H2suppers} and \ref{fig:H2slowers}.
As expected on general grounds, unlike the ground state, here $\psi_a(x)$ and 
$\psi_b(x)$ each have one node.

For the $2p\;\frac{1}{2}$ state, starting with initial parameters $B = 3.2$ eV (also presumably low), $a_0 = 0.0004$, and (fixed) $a_1 = 100.0$,
the program converged in four iterations to give final $B = 3.40086$ eV and 
$a_0 = 0.000393546$.
In fact, the $2s\;\frac{1}{2}$ and $2p\;\frac{1}{2}$ levels are degenerate \cite{Eisberg} 
as a result of an $O(4)$ symmetry hiding in the equations.
The small difference in the converged $B$'s we find here is within the 
numerical precision of our Mathematica program.
However, the $2s\;\frac{1}{2}$ and $2p\;\frac{1}{2}$ wave functions are {\it very different}.
The  $2p\;\frac{1}{2}$ wave functions are shown in Figs.\ \ref{fig:H2p12uppers} and \ref{fig:H2p12lowers}.
Note that, here also, $\psi_a$ and $\psi_b$ each have one node.

For the $2p\; \frac{3}{2}$ state, starting this time with initial parameters $B = 3.2$ eV, 
$a_0 = 0.000015$, and $a_1 = 100.0$, the program converged in four iterations to 
give final $B = 3.40144$ eV and $a_0 = 0.0000141876$.
The difference we find between the $j= \frac{3}{2}$ and $j= \frac{1}{2}$ energy levels, 
0.58 meV, is the spin-orbit splitting, an intrinsically relativistic effect.
The analytic value of this splitting is 0.453 meV \cite{BetheSal,Eisberg}
and the difference here is due to the limited machine precision used in our 
Mathematica program.
%(Perhaps we could have done better if we had forced Mathematica to carry out
%its computations with a precision greater than the default precision.)
The $2p\;\frac{3}{2}$ wave functions are shown in Figs.\ \ref{fig:H2p32uppers} and \ref{fig:H2p32lowers}.
The $\psi_a(x)$'s for the $2p\;\frac{3}{2}$ and $2p\;\frac{1}{2}$ states are {\it very similar},
but the lower component $\psi_b(x)$'s are quite different, reflecting the fact
the lower components and the difference are both entirely due to relativity.
For the $2p\;\frac{3}{2}$ state also, $\psi_a$ and $\psi_b$ each have one node.

\subsection{Hydrogen-like Atoms with $Z>1$ \label{Zgt1}}

There are no big surprises here, as Table I shows.
From Eq.\ (\ref{eq:mudef}), the exponential fall-off of the wave functions
is faster, as $\mu \sim \sqrt{B} \sim Z$, so we adjust the values of $x_{\rm match}$ and
$x_{\rm max}$ accordingly by dividing the hydrogenic values by $Z$.
Running the code for the $1s$ ground state for $V_v(r) = Z \alpha/r$ 
works well for $Z \leq 100$. 
As $Z$ grows larger, the relativistic corrections to the energy become increasingly
important. 

As $Z$ approaches $1/\alpha = 137.036$, however, the numerical solutions become 
inaccurate and, eventually, unstable.
This can be seen in the rapid increase in the value of the parameter $a_0$.
The value of $Z = 1/\alpha$ is where the Klein paradox \cite{Klein} comes into play, 
since the analytic result for this state is \cite{BjD}
\Be 
	E(1s) = m (1 - Z^2 \alpha^2)^{1/2} \label{eq:analyticEgs}
\Ee
and the eigenenergy becomes complex beyond that point.
The resolution of this paradox is, as has been well-known for a long time,
the creation of electron-positron pairs from the Dirac-Fermi sea when 
$Z > 1/\alpha$.\cite{Dombey}

\subsection{Muonic Atoms \label{muonic}}

When a negatively-charged muon slows down in matter, sometimes it is captured by the 
Coulomb potential of an atomic nucleus before it decays.
If so, it then quickly cascades down through its hydrogen-like levels to the 
$1s$ ground state, emitting x-rays along the way.
From there it then either decays or can be be captured by the nucleus, 
both of which are weak interaction processes.  
The capture process depends sensitively upon the value of the $1s$ wave function 
at the origin.

The numerical coding for muonic atom ground states for values of $Z$ of interest
goes pretty much as for the electronic atom case for such a $Z$.
The differences come from replacing the mass of the electron (0.511 MeV) with that of
the muon (105.66 MeV).
This must be done for the ``muonic Bohr radius'', $a_{B\mu} = a_B (m_e/m_\mu)$ 
= 255.8 fm, and consequently in the choices of $x_{\rm match}$ and $x_{\rm max}$
which are now proportional to $a_{B\mu}/Z$.
Also, the non-relativistic binding energy $B$ is now $Z^2 \;(m_\mu/m_e)\;$Ry.

Calculating, as above, the ground state for muonic $^{40}$Ca ($Z=20$), 
starting with $B = 1.1$ MeV and $a_0$ = 0.1,
the iterations converge to final values of $B = 1.13136$ MeV and $a_0 = 0.239462$.  
However, this initial calculation used a value of $\epsilon = 10^{-4}$ (dimensionless), 
or a cut-off at the origin of 0.000187 fm, {\it rather smaller} than the size of the 
$^{40}$Ca nucleus,\cite{Evans}
\Be 
	R_{\rm Ca} \approx r_0 A^{1/3} = 1.2 \times 40^{1/3} = 4.1 {\rm \ fm}.
\Ee

The nuclear charge is not a point charge, of course, but has a charge density 
distribution, $\rho (r)$, which is spread out over the volume of the nuclear sphere.
In fact, one of the major reasons for having studied muonic atoms in the past,
through the last $2p \rightarrow 1s$ x-ray, is for determining this charge density.\cite{mu_xray}

As a simple example, suppose the nucleus is a sphere of radius $R$ and charge $eZ$
with a charge density $\rho$ which is constant out to its surface and zero for $r>R$:
We need to find the potential seen by the negative muon for $r<R$.
If it were sitting exactly at the center of the nucleus, it would (by symmetry)
feel no force. 
That is, there is no term in the potential linear in $r$, as the force at the
origin is $-dV_{\rm nuc}/dr]_{r=0} = 0$.
This suggests that we can take $V_{\rm nuc}(r) = A + B r^2$.
The coefficient $B$ can be fixed from the requirement that the slopes 
$dV_{\rm nuc}/dr$ and $dV_{\rm Coul}/dr$ match at $r=R$.
The coefficient $A$ in turn is fixed by the requirement of continuity at $r=R$.
Thus a potential reasonably appropriate for such a muonic atom is
\Be 
	V_v(r) = \left\{\begin{array}{ll}
		\alpha Z \;(r^2 - 3 R^2)/{2 R^3} & \quad\mbox{for $r \leq R$,} 
		\\
		-\alpha Z/r                & \quad\mbox{for $r > R.$} 
		\end{array} \right.  \label{eq:unisphere}
\Ee

This potential no longer has the $1/r$ singularity at the origin, so the BCs 
for the outgoing integration for the $1s$ state are more like those in 
Eqs.\ (\ref{eq:BC34_l+1/2}) and (\ref{eq:fBC4}) for $l=0$:
\Be 
	g(\epsilon) = a_0 \; \epsilon \ , \quad \mbox{and } 
	f(\epsilon) = a_0 (1 - \tilde{E}) \; \epsilon^2/3 \ , \label{eq:muonicout}
\Ee
bearing in mind that we here are making the equations dimensionless by setting $m_\mu = 1$
and that the eigenenergy of the bound state is less than $m_\mu$.
Unlike the pure Coulomb case, here $f(r)$ will have curvature near
$r=0$, where it will vanish.

Running the numerical integrations with potential $V_v(r)$ for $^{40}$Ca gives
a binding energy $B = 1.07006$, some 60 keV smaller than the pure Coulomb result
for this nucleus.
Figure \ref{fig:CaPsis} displays the difference between the $\psi_a(x)$'s and $\psi_b(x)$'s
for the case with this nuclear charge distribution and the pure Coulomb case.

\section{Combining the Two Potentials \label{VsAndVv}}

We now consider the solution of the radial Dirac equations, Eqs.\ (\ref{eq:VvVs_geqn}) 
and (\ref{eq:VvVs_feqn}), when both potentials, the Lorentz scalar $V_s(x)$ 
and Lorentz vector $V_v(x)$, come into play.
We discuss two different situations.

\subsection{When $V_v(x)$ Is Coulombic \label{VvCoul} }

As our first case, we take the scalar potential $V_s(r)$ to be the same as in 
Eq.\ (\ref{eq:GMSSpotnl}) and the vector potential as $V_v(x) = -\alpha_s/x$.
Here $\alpha_s$ is the (strong) gluon-quark coupling constant of quantum 
chromodynamics (QCD), having a value of ${\cal{O}}(1)$ at low energies and
becoming small at high energies (asymptotic freedom).

The modifications to the program are minor.
The two BC's at $x_{\rm max}$ are the same as in the pure $V_s$ case, Eq.\ (\ref{eq:BC12}).
The BC's at $x=\epsilon$ are slightly modified from that of 
Eq.\ (\ref{eq:BC34_l+1/2}) because here it is $V_v$ that dominates.
For $l = 0$,
\Be 
	g(\epsilon) = a_0\;\epsilon, \quad
	f(\epsilon) = -a_0\;(V_v(\epsilon)/3)\;\epsilon^2 = a_0\;\alpha_s\;\epsilon/3
	\ , \label{eq:VvVsBC4}
\Ee
as in the hydrogen atom case, but with $\alpha_s$ generally much larger than $\alpha$.
 
Running the code for $\alpha_s = 0.4$ gives the wave functions $\psi_a$ and  $\psi_b$
shown in Fig.\ \ref{fig:VsVv1sab_04}.
For comparison, the wave functions for case of $\alpha_s = 0$ (i.e., only using 
the scalar potential) are also shown.
The eigenenergies for these two cases are 0.306 GeV ($\alpha_s = 0$) and
0.251 GeV ($\alpha_s = 0.4$).
That is, adding the Coulomb attraction lowers the ground state energy.

The sharp rise in $\psi_a(x)$ and rapid falloff in $\psi_b(x)$ near the origin is another
effect of the point Coulomb attraction.
This behavior is unphysical, as the motions of the quark and anti-quark in the $q\bar{q}$
meson will smear out the point Coulomb potential somewhat.
One might think that a Coulomb divergence remains in the relative coordinate,
but this is a non-relativistic prejudice.  
Relativistic retardation effects and pair creation and annihilation both act to ameliorate 
this divergence, significantly so in this case, as $\alpha_s \gg \alpha$.
This will give a potential $V_v(x)$ that is similar to the potential used to describe the
nuclear size effect in muonic atoms, Eq.\ (\ref{eq:unisphere}).
We will come back to such a calculation presently.

Before doing that, however, we present the results of running the massless quarks code
for a variety of choices for $\alpha_s$.
The $1s$ ground state energies and the inital slope parameters $a_0$
are shown in the second and third columns of Table II.
As the Coulomb attraction increases, the eigenenergy falls off fairly rapidly
and eventually goes through zero at a value near $\alpha_s = 1$.
At this point, like the Klein paradox situation discussed in 
Sec.\ {\ref{Zgt1}, it is possible to produce $q\bar{q}$ pairs at liberty.
In fact,  what is called a ``quark condensate'' forms, a modification of the
vacuum state.\cite{GOR}

To take into account the smearing of the point Coulomb potential by the motions of the
quarks in the meson, we again use the potential $V_v(x)$ given in Eq.\ (\ref{eq:unisphere}),
but with $\alpha$ replaced by $\alpha_s$ and $R$ taken as one half of the electromagnetic 
radius of the pion, i.e., $R = \frac{1}{2}\times 0.672 = 0.336$ fm.\cite{PDG}\ 
We choose this fraction on the basis that the root-mean-square electromagnetic radius
represents the average separation of the quark and antiquark, but since they are oppositely
correlated, each should be at about half of this distance from the center of mass of 
the meson.

The boundary conditions (BC's) at the origin for this smeared Coulomb potential 
(for angular momentum $j = l + \frac{1}{2}$) now become
\Be 
	g(\epsilon) = a_0\;\epsilon^{l+1}, \quad
	f(\epsilon) = -a_0\;(\tilde{E}- V_v(\epsilon) - V_s(\epsilon) - \tilde{m}) \;
		\epsilon^{l+2}/(2l+3) \ . \label{eq:VvVsBC4smeared}
\Ee
Running the program with $\alpha_s = 0.4$ and the above $R$ 
gives curves for $\psi_a(x)$ and $-\psi_b(x)$ with similar upturns and falloffs near the 
origin as in Fig.\ \ref{fig:VsVv1sab_04}, but less sharply so.
One should be aware that, while interesting, these differences in the $\psi_{a,b}(x)$
near the origin make only minor differences in expectation values of operators because of
the factor of $x^2$ in the integration over $d^3x$; see, e.g., the normalization condition,
Eq.\ (\ref{eq:normzn}).

For this smeared potential, the eigenenergies and final $a_0$ parameters for various 
values of $\alpha_s$ are also tabulated in Table II.
As expected from the above remark, the energy expectation value is basically unchanged 
from the point Coulomb value until one reaches the point where the quark condensate 
takes over.
However, the initial slope parameters $a_0$ do differ from those of the point Coulomb
case, being less pathological as $E$ goes through zero.

\subsection{$V_v$ as the Cornell Potential and the Spin-orbit Force. \label{CornellPot}}

In this subsection we will assume both potentials $V_s(x)$ and $V_v(x)$ are linearly 
confining and have the same slopes (i.e., have the same string tension $\kappa^2$).
We assume the two potentials are displaced from each other with $V_s(x)$ lying above 
$V_v(x)$, as in Fig.\ \ref{fig:VsVvplot}.
That is, we assume
\Be 
	V_s(x) = x + x_s, \quad V_v(x) = -\alpha_s/x + x - x_v\ , \label{eq:VsVvCornell}
\Ee
or, more appropriately, the smeared version of $V_v(x)$ as described above in 
Sec.\ \ref{VvCoul}.
Here $x_s$ and $x_v$ are parameters to be adjusted to get some desired eigenenergy.
They give rise to a separation between the two asymptotic potentials of $x_d = x_s + x_v$.

The reason for giving the potentials the same asymptotic slopes is to check that 
the spin-orbit interaction disappears as claimed in Ref.\ \cite{PGG}.
In doing so, we have changed the earlier offset of $-x_0$ in $V_s$ to $+x_s$ since, 
in GMSS (where only a Lorentz scalar potential was used), the negative offset $-x_0$  
represented a reasonable approximation the one-gluon-exchange attraction, 
roughly an average of the two potentials as shown in the figure.

Now, however, the BC's at large distances are changed from those in Eq.\ (\ref{eq:BC12}),
since, in the equation for $g(x)$, Eq.\ (\ref{eq:VvVs_geqn}),
the difference of the dimensionless $V_s(x)$ and $V_v(x) \rightarrow x$ has its 
$x$-dependence cancel.
We need to solve the large-$x$ equations
\Be 
	g\;'(x) - (\tilde{E} + x_d + \tilde{m})\; f(x) = 0, \quad 
	f\;'(x) - 2x \; g(x) = 0 \ , \label{eq:preAiry}
\Ee
from which
\Be 
	g\;''(x) - 2(\tilde{E} + x_d + \tilde{m})\; x \;g(x) = 0 \ . \label{eq:thisAiry}
\Ee
This is a minor modification of the differential equation for the Airy functions.\cite{AbStAiry}\ \ 
Since we require decaying solutions at our starting point $x_{\rm max}$ for the 
inwards integrations, we discard the runaway $Bi$ solution for $g(x)$.
The BC's at large $x=x_{\rm max}$ are then 
\Bea 
	g(x_{\rm max}) &=& a_1 \;Ai\;[2^{1/3} (\tilde{E} + x_d + \tilde{m})^{1/3} x_{\rm max}\;] 
		\ , \nonumber \\
	f(x_{\rm max}) &=& a_1 \left[\frac{2} {(\tilde{E} + x_d +  \tilde{m})^2}\right]^{1/3}
		\!\!Ai\;'\;[\;2^{1/3} (\tilde{E} + x_d +  \tilde{m})^{1/3} x_{\rm max}\;] \ . \label{eq:AiryBCs}
\Eea
Despite appearances, $g(x_{\rm max})$ and $f(x_{\rm max})$ still
start off with the opposite slopes, as in Eq.\ (\ref{eq:BC12}), since, for
$t > 0$, $Ai(t) > 0$ and $Ai\;'(t) < 0$.

For calculations, we simplified the displacement between the two potentials by
setting $x_v = 0$, giving us one less parameter to worry about.
The first wave function we want to compute is the ground state for the massless
quark case, $m_q = 0$, starting out with $\alpha_s = 0$ and then adjusting both
$x_d = x_s$ and  $\alpha_s$ until we get the desired eigenenergy, noted above in
Eq.\ (\ref{eq:m_ud}).

It turns out that to do so is trickier than in the $V_s$-only case, presumably because of
the cancellation between $V_s(x)$ and $V_v(x)$ in the differential equation for $g\;'(x)$.
Our first iterative search with $\alpha_s = 0$ found a highly excited state with 
$\tilde{E} = 7.975$, {\it not} the desired 0.7263, for which $\psi_a(x)$ had six nodes!
After some hand-searching on $x_s$, we found we {\it could} get a ground state solution,
i.e., a $\psi_a(x)$ with no nodes, looking like the $b$-quark $\psi_a(x)$ of 
Fig.\ \ref{fig:upperWFs}.
This solution started with $x_s = 4.0$ (very large!) and converged, in six iterations,
to $\tilde{E} = 5.735$ (about eight times too large).

From this $\alpha_s = 0$ solution we were able to march down, in small steps in $x_s$ 
(still with $\alpha_s = 0$), to $x_s = 0.20$, with the energy now smaller, 
$\tilde{E} = 2.779$, but still a factor of four too large.
(For smaller values of $x_s$ the iterative procedure would not converge, with the matrix
of Eq.\ (\ref{eq:2x2partials}) becoming singular.)
The wave function $\psi_a(x)$ for this solution was broader than that for the $x_s = 4.0$
case, and the $\psi_b(x)$ was relatively larger in comparison with $\psi_a$.

At this point we began marching the value of $\alpha_s$ in small steps up from zero, 
i.e., we began turning on the Coulomb attraction.
As expected this attraction lowers the energy, and we were able to achieve the desired
massless quark eigenenergy of 0.306 GeV with the following values:
\Be 
	x_s = 0.20, \quad \alpha_s = 1.103, \quad \tilde{E} = 0.7263 \ . \label{eq:VsVv1s_ud}
\Ee
The Long Marches did achieve the factor of eight reduction in the ground state eigenenergy. 
The wave functions for this massless quark solution are shown in Fig.\ \ref{fig:VsVv1s_ud}.
Note that $\psi_a(x)$ here begins to resemble the hydrogen-atom ground state $\psi_a(x)$ 
depicted by the dashed curve in Fig.\ \ref{fig:Huppers}, and it is much narrower than 
that shown in Fig.\ \ref{fig:finalVs1sPsis} .

The ground state solution says nothing about the PGG claim that, 
for $V_s(x)$ and $V_v(x)$ having equal slopes, the spin-orbit interaction vanishes.  
To check this we need to find, at a minimum, the eigenenergies of the
$2p\;\frac{1}{2}$ and $2p\;\frac{3}{2}$ excited states. 
For the same values of $x_s$ and $\alpha_s$ as in Eq.\ (\ref{eq:VsVv1s_ud}) and 
after some searching for good starting values for the iterative procedure,
we obtained the three $n=2$ wave functions displayed in Figs.\ \ref{fig:VsVv2s_ud},
\ref{fig:VsVv2p1by2_ud}, and \ref{fig:VsVv2p3by2_ud}.
The wave functions here are narrower and more hydrogen-like 
than in the $V_s$-only case (compare with Figs.\ \ref{fig:finalVs2s},
\ref{fig:finalVs2p3by2}, and \ref{fig:finalVs2p1by2}).
The eigenenergies, in GeV, for these states are
\Be 
	E_{2s} = 1.020, \quad E_{2p\;\frac{1}{2}} = 0.674, \quad E_{2p\;\frac{3}{2}} = 1.008 \ .
	\label{eq:VsVvn=2Es}
\Ee
The $2p\;\frac{3}{2}$ state is nearly degenerate with the $2s$ state,
some 200 MeV higher than in Sec.\ \ref{VsExciteds}, but this degeneracy is 
probably accidental.
The real surprise here is the {\it huge} spin-orbit splitting between the 
$2p\;\frac{1}{2}$ and $2p\;\frac{3}{2}$ states, not what we expected from 
the claim made by PGG.\cite{PGG}

This difference, it turns out, is due to the presence of 
the Coulomb attraction term in $V_v(x)$.  
If we set $\alpha_s = 0$, then (again for $x_s=0.2$) we find the following
$n=2$ eigenenergies (in GeV):
\Be 
	E_{2s} = 1.757, \quad E_{2p\;\frac{1}{2}} = 1.523, 	\quad E_{2p\;\frac{3}{2}} = 1.523 \ .
	\label{eq:VsVvn=2EsNoCoul}
\Ee
The wave functions in this case are similar to those in Figs.\ \ref{fig:VsVv2s_ud},
\ref{fig:VsVv2p1by2_ud}, and \ref{fig:VsVv2p3by2_ud} but are somewhat 
broader and less peaked near the origin.
These results show the expected PGG symmetry (zero spin-orbit splitting).
Note, incidentally, that the $2s$ state is higher in energy than the $2p$ states,
as in the experimental charmonium spectrum.\cite{PDG}
One does conclude, however, that having a significant Coulomb contribution 
in $V_v(x)$ will destroy the PGG symmetry.

%{\bf NEEDS A CONCLUSION SECTION}
\section{Summary}

In this paper we have discussed, pedagogically, three increasingly more 
sensitive cases for solving the radial Dirac equations numerically:
\begin{enumerate}
\item
    A Lorentz scalar potential, $V_s$, which is linearly confining for quarks.  The 
    case of massless quarks is necessarily relativistic.  As the mass of the quark
    increases, the wave function solutions become more and more n0n-relativistic.
\item   
    A Lorentz vector potential, $V_v$, whose time-like component is a 
    Coulomb potential.  We first treated, numerically, the hydrogen atom, which is 
    basically non-­relativistic problem. 
    We do, however, obtain the relativistic spin-­orbit splitting between the 
    $2p \frac{1}{2}$ and $2p \frac{3}{2}$ states (at the edge of our machine precision).
\item    
    The combination of scalar and vector potentials, aiming at a relativistic
    quark model of mesons.
    Of particular interest is when we do (and do not) get a relativistic
    spin-­orbit splitting.
\end{enumerate}

We thank L.\ J.\ Curtis, J.\ L.\ Friar, C.\ J.\ Fontes, M.\ D.\ Scadron, and R. Sharma for 
discussions and comments on this work.

%------------------------------ 
%\pagebreak

\pagebreak

\begin{figure} %Fig. 1
\includegraphics[width=.7\textwidth, height=0.45\textwidth]{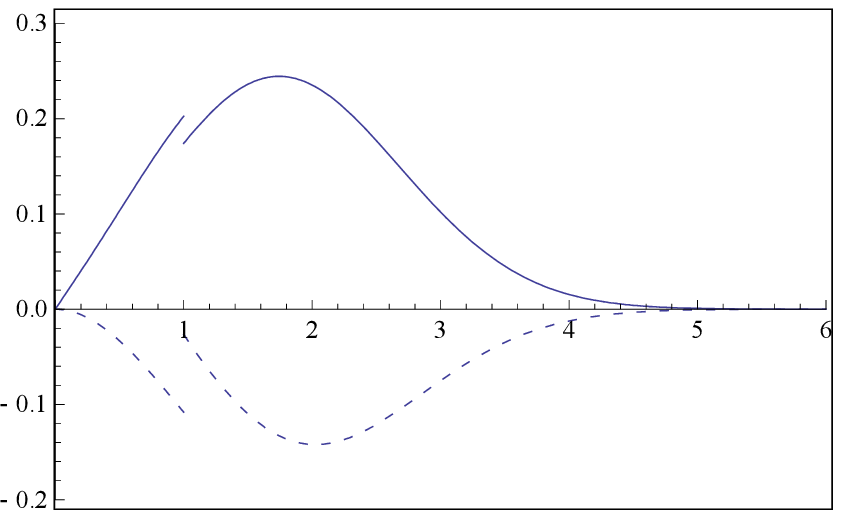}
\caption{The first-pass $1s$ ground state radial wave functions $g(x)$ (solid curve) 
and $f(x)$ (dashed curve) for massless quarks
in the GMSS linear potential, Eq.\ (\ref{eq:GMSSpotnl}), given the initial guess for parameters $\tilde{E}$ and $a_0$ of Eq.\ (\ref{eq:params1sVs}).
\label{fig:initialVs1s}}
\end{figure} 

\begin{figure} %Fig. 2
\includegraphics[width=.7\textwidth,height=0.45\textwidth]{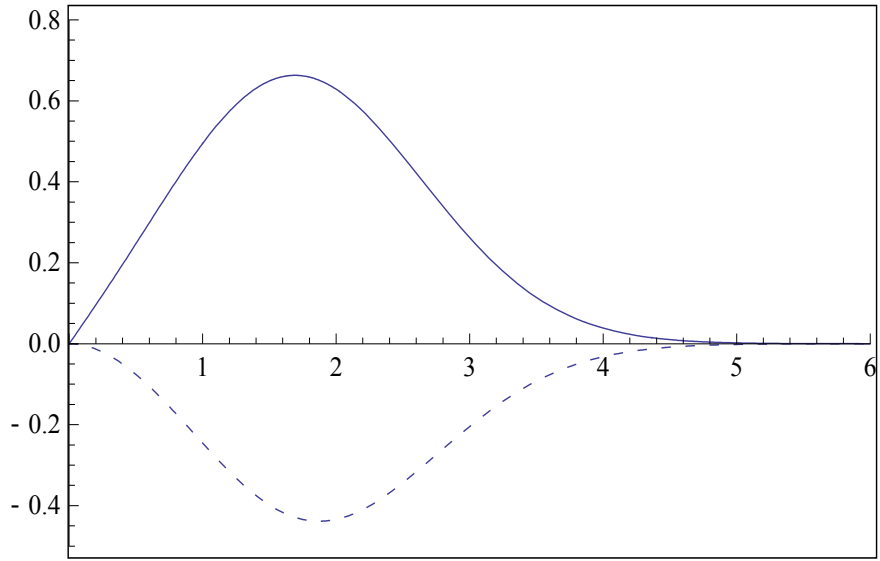}
\caption{The normalized $1s$ ground state radial wave functions $g(x)$ and $f(x)$ 
for massless quarks in the GMSS potential, after iterations have converged to 
final parameters $\tilde{E}$ and $a_0$. 
\label{fig:finalVs1s}}
\end{figure} 

\begin{figure} %Fig. 3
\includegraphics[width=.7\textwidth,height=0.45\textwidth]{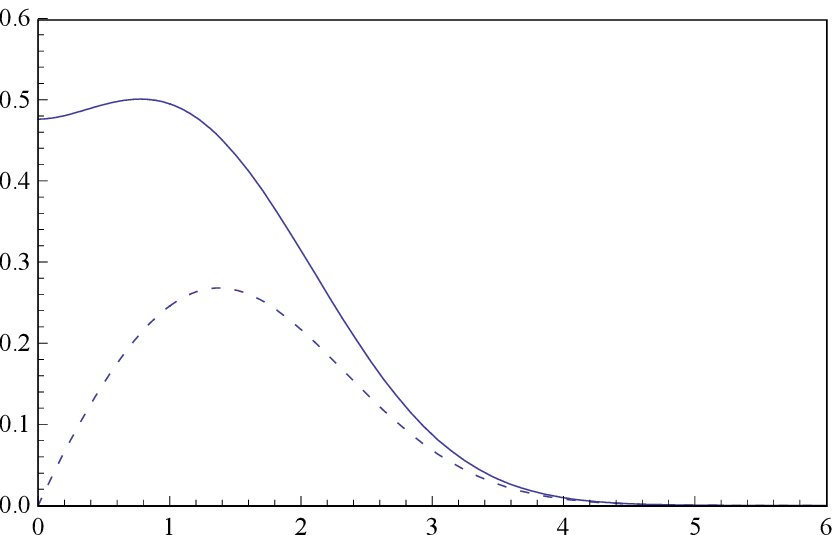}
\caption{The normalized $1s$ ground state radial wave functions $\psi_a(x)$ 
(solid curve) and $-\psi_b(x)$ (dashed curve) for massless quarks
in the GMSS linear potential for fitting $q\bar{q}$ mesons. 
\label{fig:finalVs1sPsis}}
\end{figure}

\begin{figure} %Fig. 4
\includegraphics[width=.7\textwidth,height=0.45\textwidth]{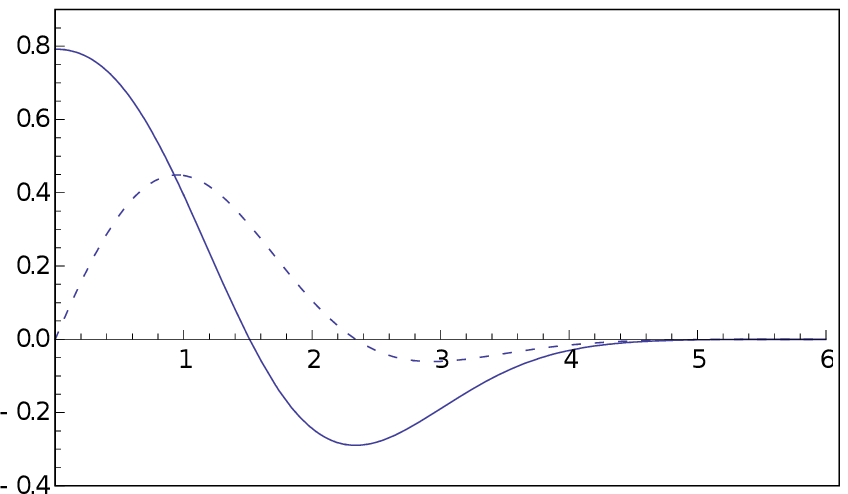}
\caption{The normalized $2s$ excited state radial wave functions $\psi_a(x)$ 
(solid curve) and $-\psi_b(x)$ (dashed curve) for massless quarks
in the GMSS linear potential for fitting $q\bar{q}$ mesons. 
\label{fig:finalVs2s}}
\end{figure}

\begin{figure} %Fig. 5
\includegraphics[width=.7\textwidth,height=0.45\textwidth]{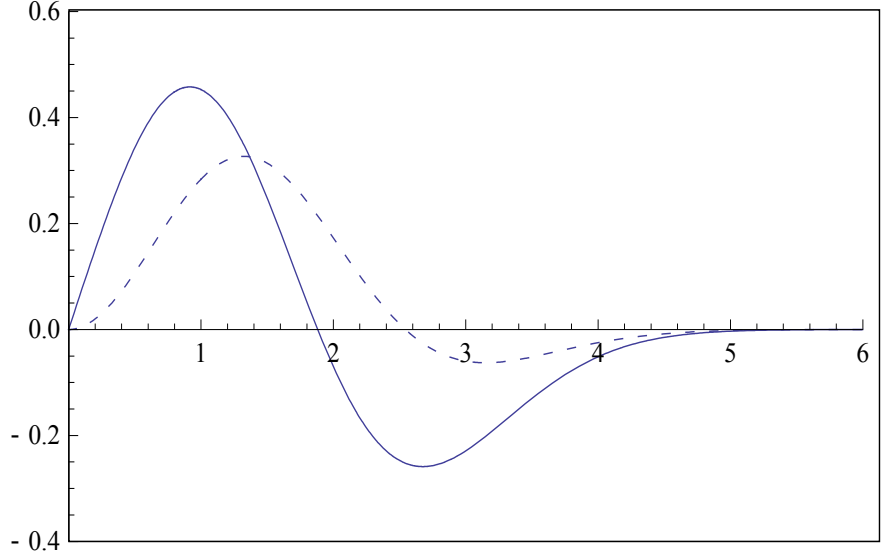}
\caption{The normalized $2p \; \frac{3}{2}$ excited state $\psi_a(x)$  
(solid curve) and $-\psi_b(x)$ (dashed curve) for massless quarks
in the GMSS potential. \label{fig:finalVs2p3by2}}
\end{figure}

\begin{figure} %Fig. 6
\includegraphics[width=.7\textwidth,height=0.45\textwidth]{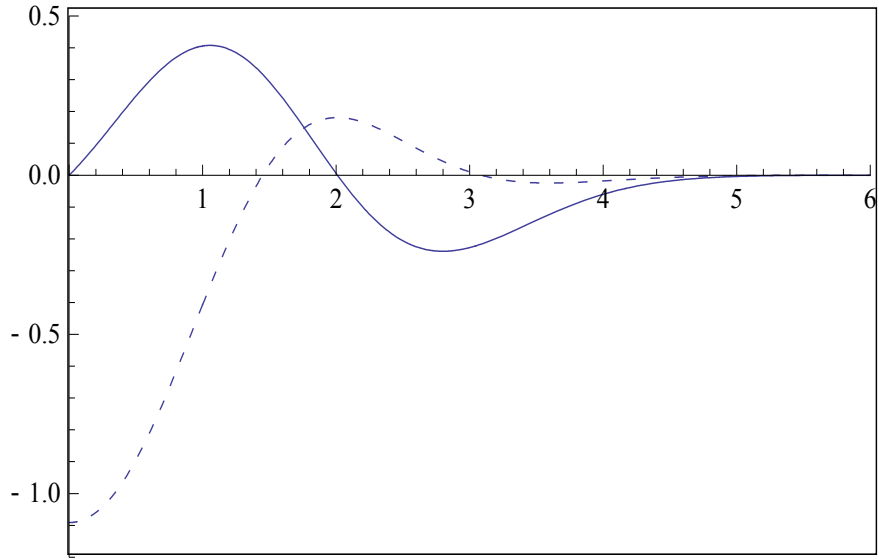}
\caption{The normalized $2p \; \frac{1}{2}$ excited state $\psi_a(x)$ 
(solid curve) and $-\psi_b(x)$ (dashed curve) for massless quarks
in the GMSS potential. 
Note the considerable differences between these wave functions and those in Figs.\ 4 or 5.
\label{fig:finalVs2p1by2}}
\end{figure}

\begin{figure} %Fig. 7 uppers
\includegraphics[width=.7\textwidth,height=0.45\textwidth]{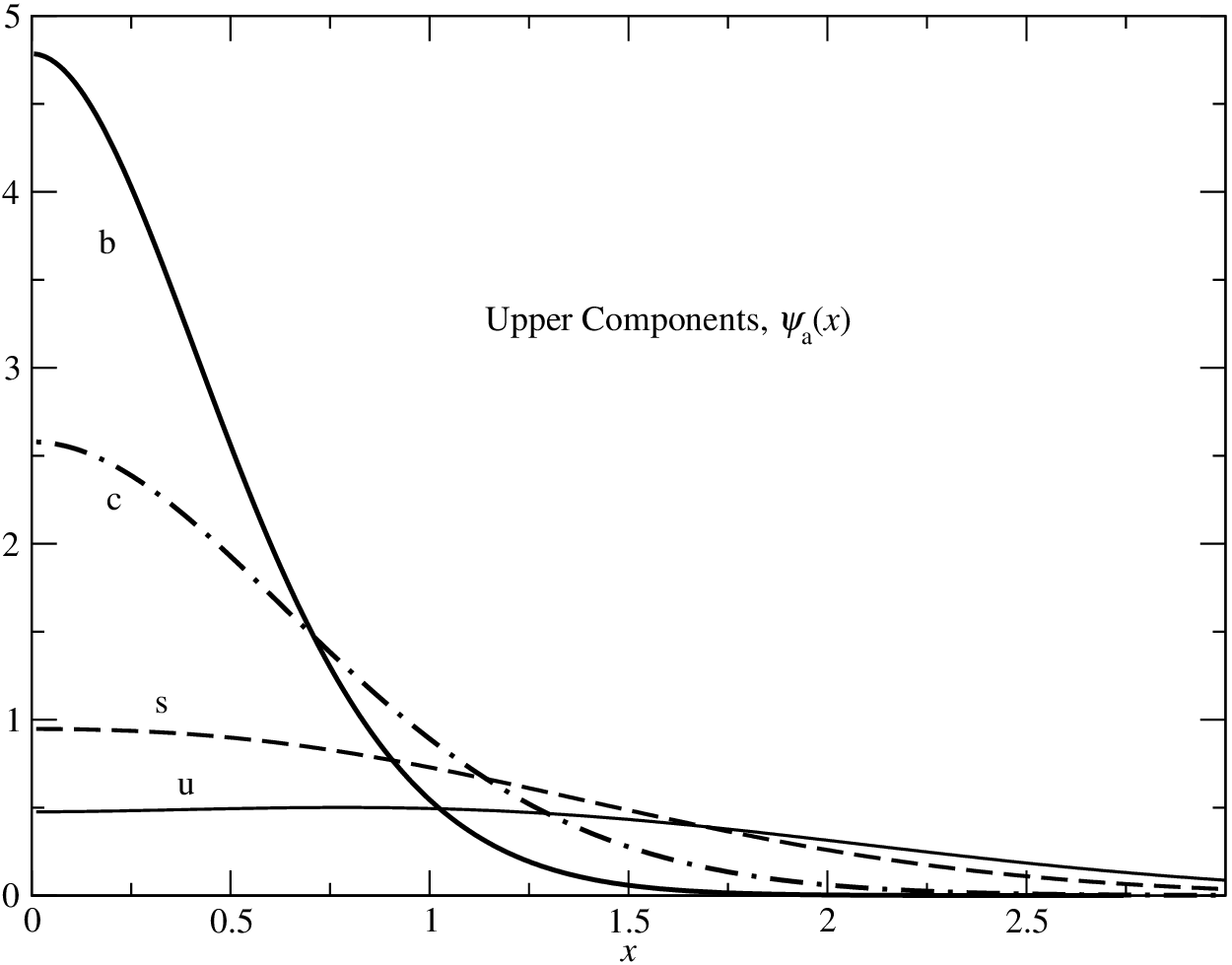}
\caption{The upper component radial wave functions, 
$\psi_a(x)$, for massless $u$ and massive $s$, $c$, and $b$ quarks. 
\label{fig:upperWFs}}\vspace{0.5in}
\end{figure} 

\begin{figure} %Fig. 8 lowers
%\rotatebox{-90}
{\includegraphics[width=.7\textwidth,height=0.45\textwidth]{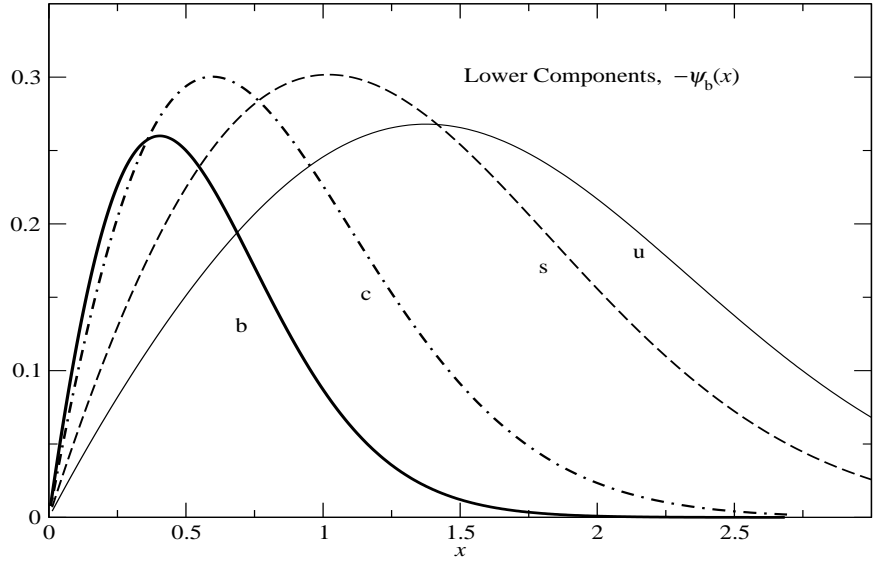}}
\caption{The lower component radial wave functions, 
$-\psi_b(x)$, for massless $u$ and massive $s$, $c$, and $b$ quarks. 
\label{fig:lowerWFs}}
\end{figure} 

\begin{figure} %Fig. 9
\includegraphics[width=.7\textwidth,height=0.45\textwidth]{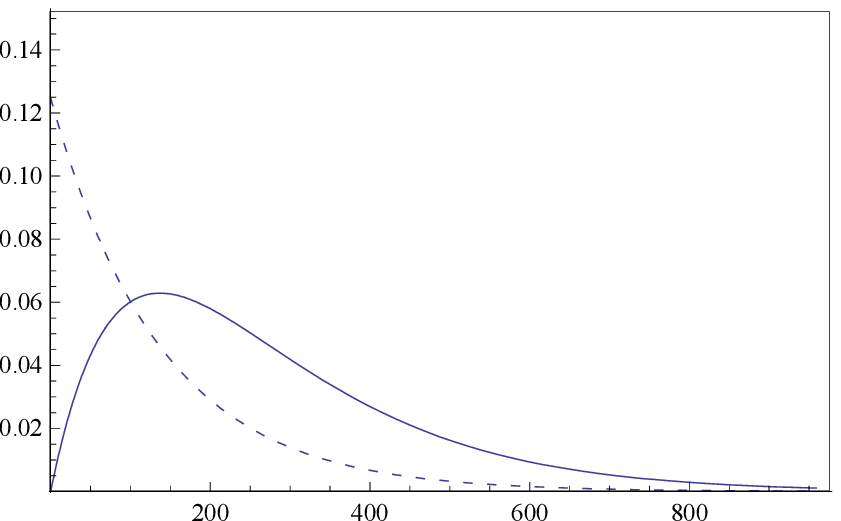}
\caption{The normalized hydrogen atom ground state upper component wave functions 
$g(x)$ (solid curve) and $\psi_a(x) = g(x)/x$ (dashed curve).
\label{fig:Huppers}}
\end{figure}

\begin{figure} %Fig. 10
\includegraphics[width=.7\textwidth,height=0.45\textwidth]{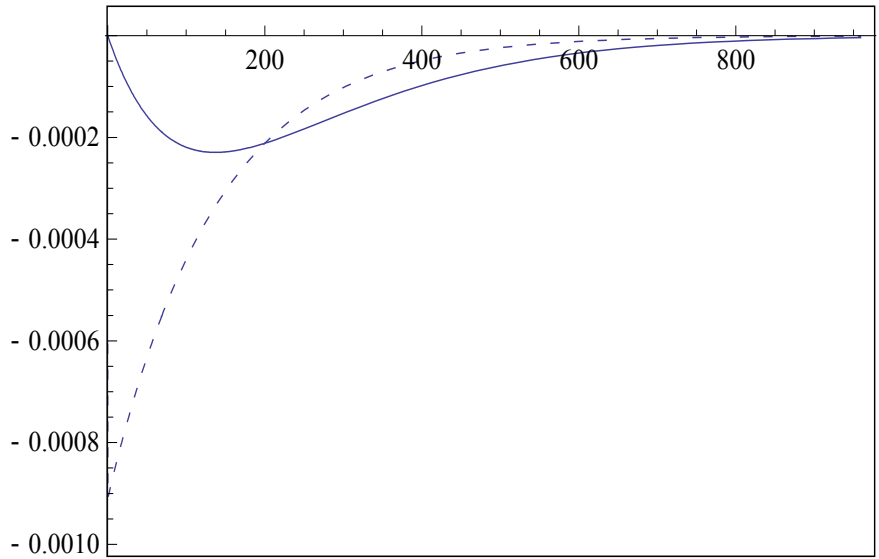}
\caption{The normalized hydrogen atom ground state lower component wave functions 
$f(x)$ (solid curve) and $\psi_b(x) = f(x)/x$ (dashed curve).  Note the smallness
of these wave functions compared to those for the upper component.
\label{fig:Hlowers}}
\end{figure}

\begin{figure} %Fig. 11
\includegraphics[width=.7\textwidth,height=0.45\textwidth]{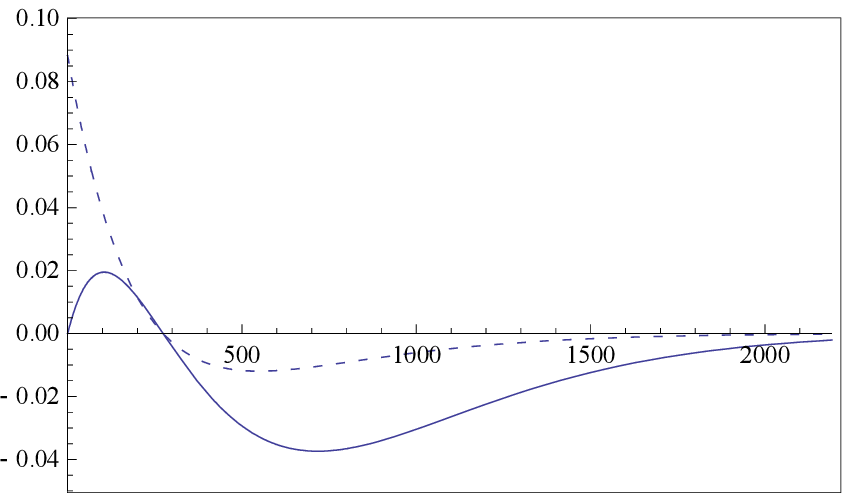}
\caption{The normalized hydrogen atom $2s$ state upper component wave function %s 
$g(x)$ (solid curve) and $\psi_a(x) = g(x)/x$ (dashed curve).
\label{fig:H2suppers}}
\end{figure}

\begin{figure} %Fig. 12
\includegraphics[width=.7\textwidth,height=0.45\textwidth]{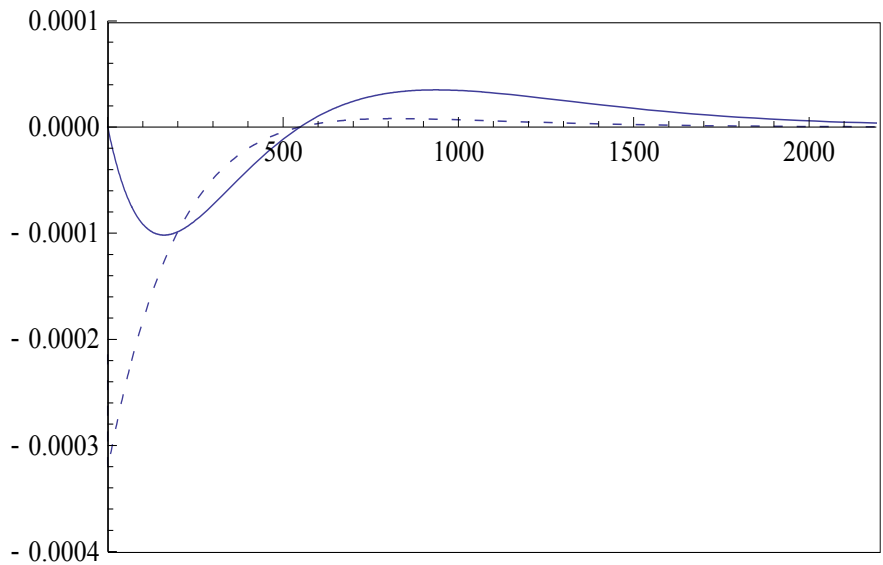}
\caption{The normalized hydrogen atom $2s$ state lower component wave function %s 
$f(x)$ (solid curve) and $\psi_b(x) = f(x)/x$ (dashed curve).
\label{fig:H2slowers}}
\end{figure}

\begin{figure} %Fig. 13
\includegraphics[width=.7\textwidth,height=0.45\textwidth]{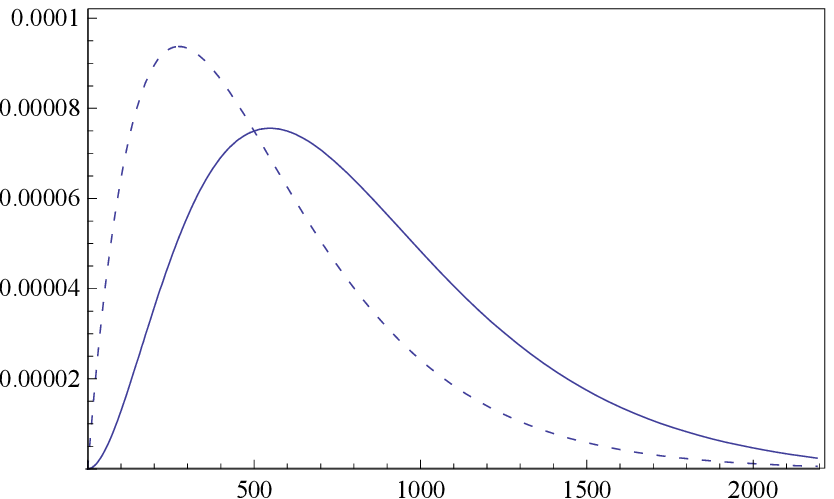}
\caption{The normalized hydrogen atom $2p\; \frac{1}{2}$ state upper component wave functions 
$0.002\times g(x)$ (solid curve) and $\psi_a(x) = g(x)/x$ (dashed curve).
\label{fig:H2p12uppers}}
\end{figure}

\begin{figure} %Fig. 14
\includegraphics[width=.7\textwidth,height=0.45\textwidth]{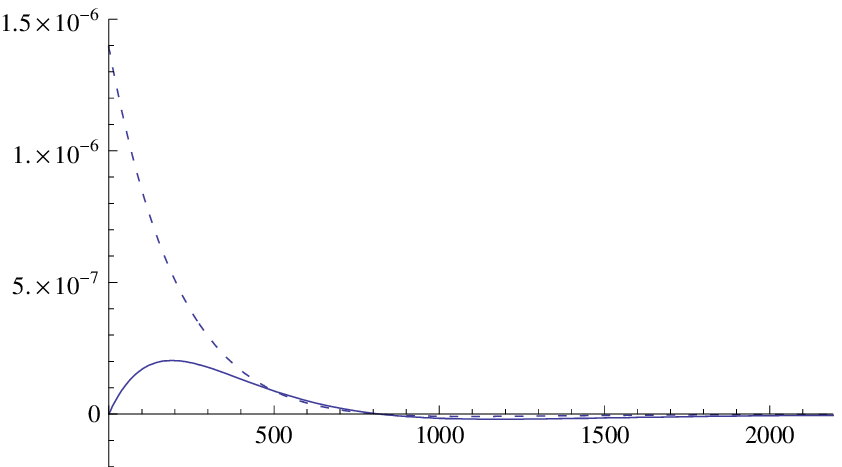}
\caption{The normalized hydrogen atom $2p\; \frac{1}{2}$ state lower component wave functions
$0.002\times f(x)$ (solid curve) and $\psi_b(x) = f(x)/x$ (dashed curve).
\label{fig:H2p12lowers}}
\end{figure}

\begin{figure} %Fig. 15
\includegraphics[width=.7\textwidth,height=0.45\textwidth]{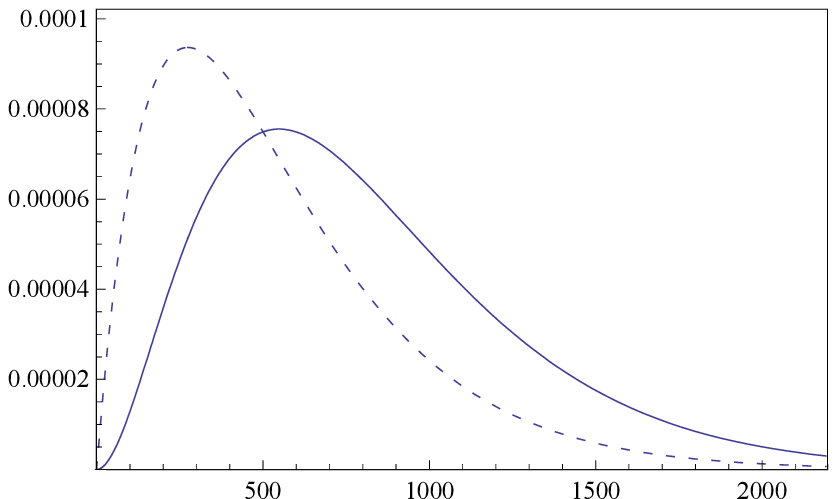}
\caption{The normalized hydrogen atom $2p\; \frac{3}{2}$ state upper component wave functions 
$0.002 \times g(x)$ (solid curve) and $\psi_a(x) = g(x)/x$ (dashed curve). 
\label{fig:H2p32uppers}}
\end{figure}

\begin{figure} %Fig. 16
\includegraphics[width=.7\textwidth,height=0.45\textwidth]{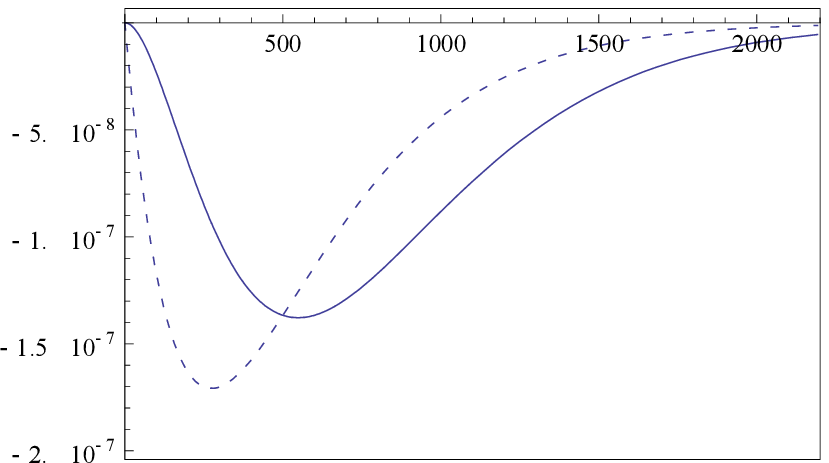}
\caption{The normalized hydrogen atom $2p\; \frac{3}{2}$ state lower component wave functions
$0.002\times f(x)$ (solid curve) and $\psi_b(x) = f(x)/x$ (dashed curve).
\label{fig:H2p32lowers}}
\end{figure}

\begin{figure} %Fig. 17
\includegraphics[width=0.8\textwidth,height=0.5\textwidth]{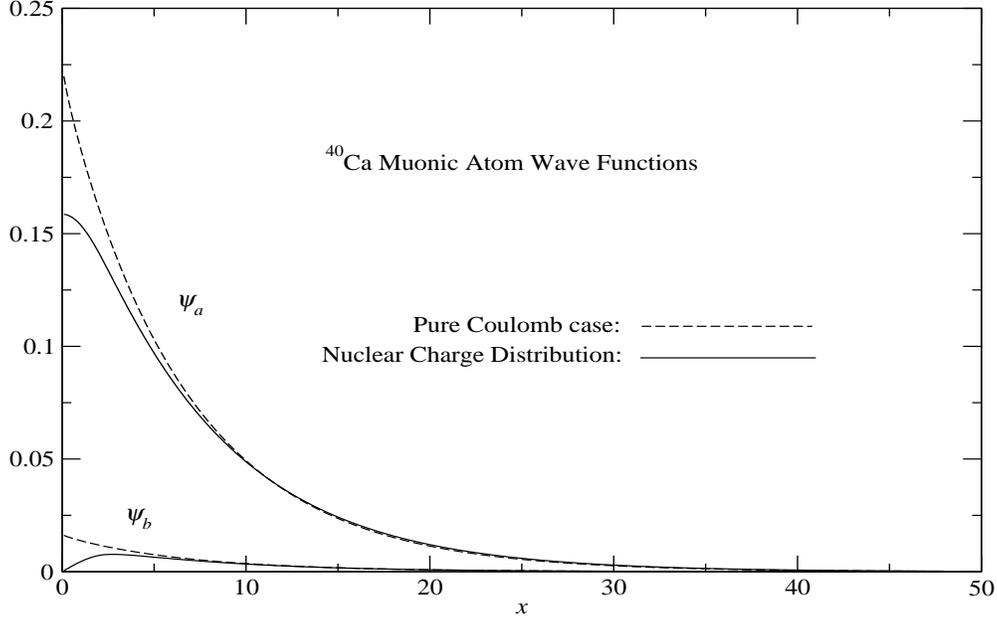}
\caption{The {\it unnormalized} $1s$ muonic atom wave functions for $^{40}$Ca (solid
curves) compared with those for a pure point Coulomb potential with charge $Z=40$
(dashed curves).
\label{fig:CaPsis}}
\end{figure}

\begin{figure} %Fig. 18
\includegraphics[width=0.8\textwidth,height=0.5\textwidth]{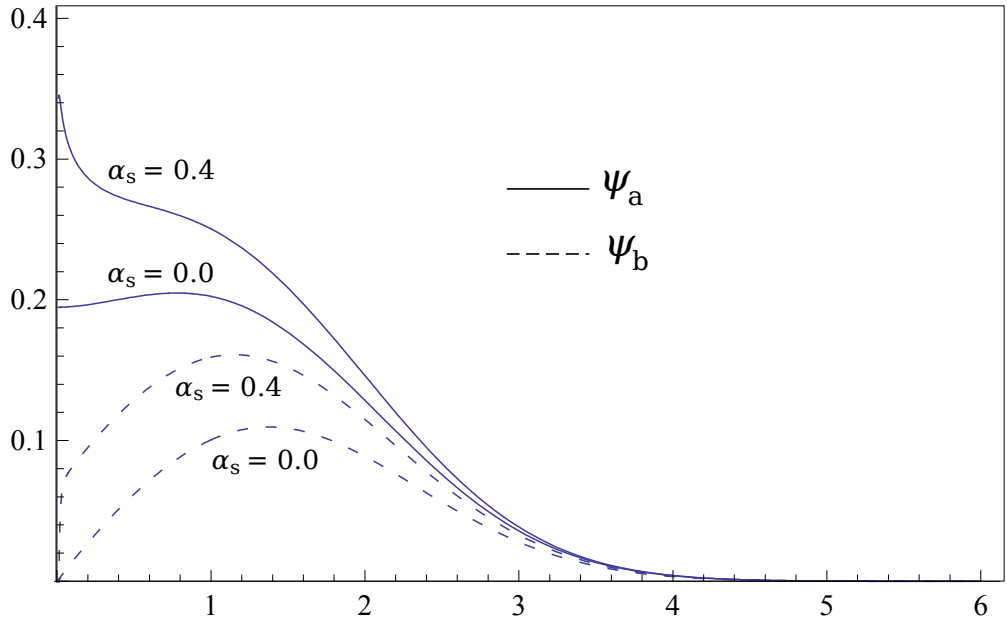}
\caption{The {\it unnormalized} $1s$ wave functions for massless quarks,
$\psi_a$ (upper solid curve) and $\psi_b$ (upper dashed curve), for both potentials 
$V_v = -\alpha_s/r$ and $V_s$ of Eq.\ (\ref{eq:GMSSpotnl}).
The lower curves are for $\alpha_s = 0$, and are the same as in Fig.\ \ref{fig:finalVs1sPsis}.
The value of $\alpha_s$ used for the upper curves was 0.4.
\label{fig:VsVv1sab_04}}
\end{figure}

\begin{figure} %Fig. 19 
\includegraphics[width=0.7\textwidth,height=0.5\textwidth]{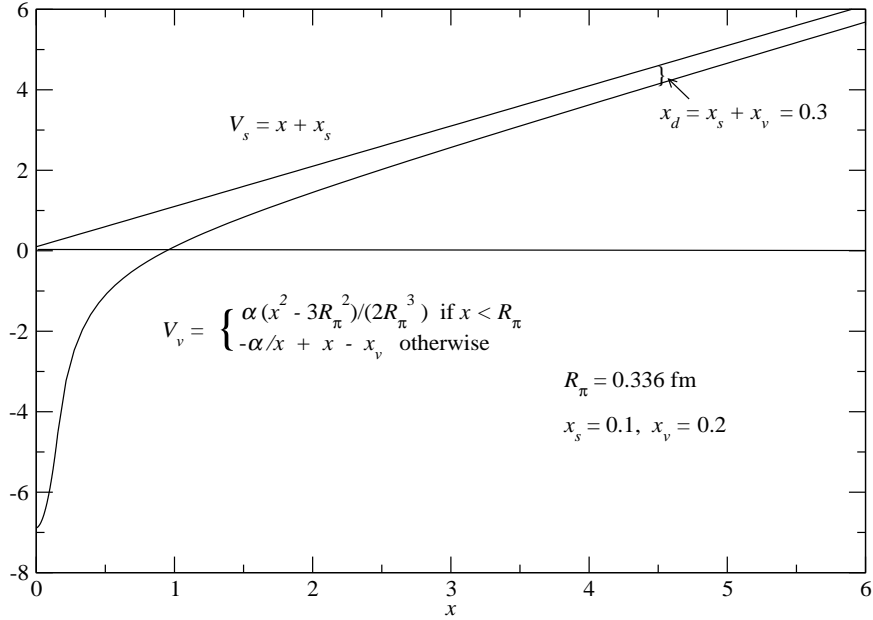}
\caption{Example of having the (dimensionless) Lorentz scalar $V_s(x)$ and 
(smeared) vector $V_v(x)$ potentials with the same string tension 
(i.e., the same slopes at large $x$.)}
\label{fig:VsVvplot}%\pagebreak
\end{figure}

\begin{figure} %Fig. 20  
\includegraphics[width=0.7\textwidth, height=0.5\textwidth]{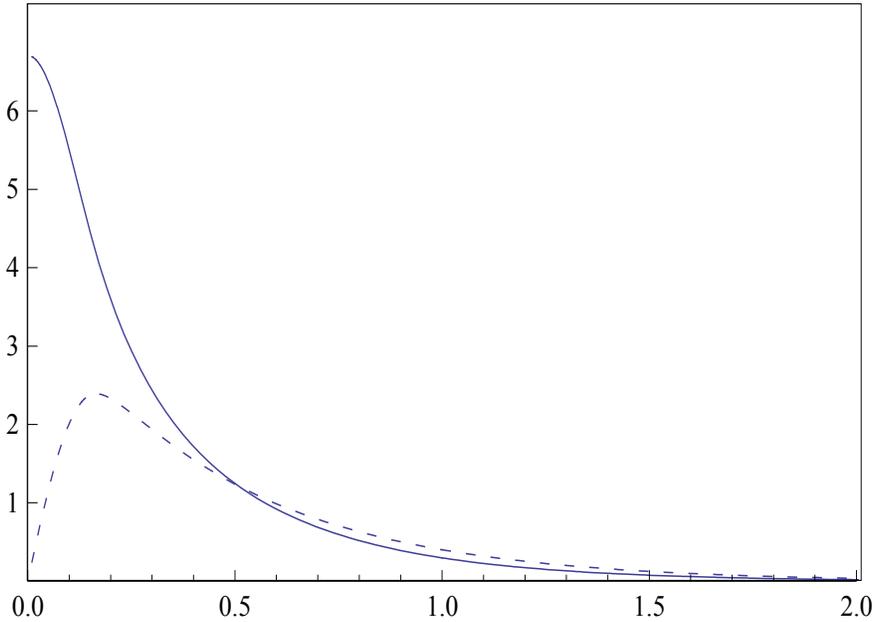}
\caption{The normalized ground state wave functions $\psi_a(x)$ (solid curve)
and $-\psi_b(x)$ (dashed curve)
when $V_s(x)$ and $V_v(x)$ have equal asymptotic slopes.  
This is the solution for the 
massless quark case ($u$ and $d$ quarks) using the parameters given in 
Eq.\ (\ref{eq:VsVv1s_ud}).  Note that the wave functions are very narrow, 
having decayed away by $x = 2$.
Compare this solution for the massless quark case shown 
in Fig. \ref{fig:finalVs1sPsis}.}
\label{fig:VsVv1s_ud}\pagebreak
\end{figure}

\begin{figure} %Fig. 21  
\includegraphics[width=0.7\textwidth, height=0.4\textwidth]{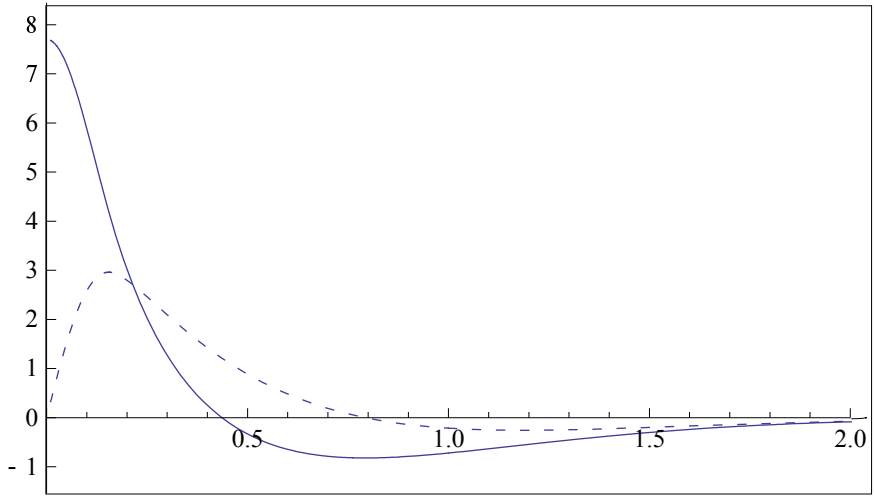}
\caption{The normalized first excited $2s$ wave functions $\psi_a(x)$ and $-\psi_b(x)$ 
for the case when $V_s(x)$ and $V_v(x)$ have equal asymptotic slopes.
Compare this solution for the massless quark case ($u$ and $d$ quarks) shown 
in Fig. \ref{fig:finalVs2s}.
Compare also with the $2s$ hydrogen-atom wave functions, $\psi_{a,b}(x)$  
in Figs.\ \ref{fig:H2suppers} and \ref{fig:H2slowers} (dashed curves), 
noting that the smeared Coulomb 
attraction in $V_v$ now forces $\psi_b(0)$ to be 0 instead of a finite value.}
\label{fig:VsVv2s_ud}
\end{figure}

\begin{figure} %Fig. 22  fig:finalVsVv2p3by2_ud
\includegraphics[width=0.7\textwidth, height=0.5\textwidth]{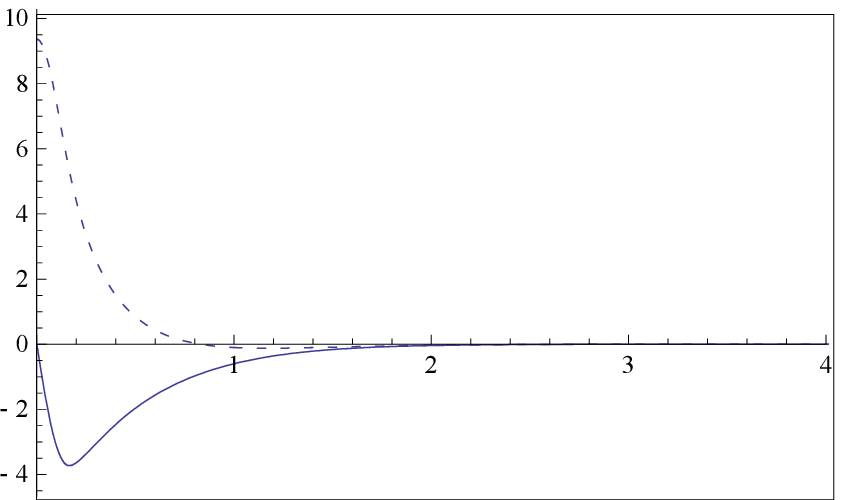}
\caption{The normalized first excited $2p\;\frac{1}{2}$ wave functions $\psi_a(x)$ and -$\psi_b(x)$ 
for the case when $V_s(x)$ and $V_v(x)$ have equal asymptotic slopes.
Compare this solution for the massless quark case ($u$ and $d$ quarks) shown 
in Fig. \ref{fig:finalVs2p1by2}.
Compare also with the $2p\;\frac{1}{2}$ hydrogen-atom wave functions, $\psi_{a,b}(x)$ 
in Figs.\ \ref{fig:H2p12uppers} and \ref{fig:H2p12lowers} (dashed curves).}
\label{fig:VsVv2p1by2_ud}
\end{figure}

\begin{figure} %Fig. 23  
\includegraphics[width=0.7\textwidth, height=0.4\textwidth]{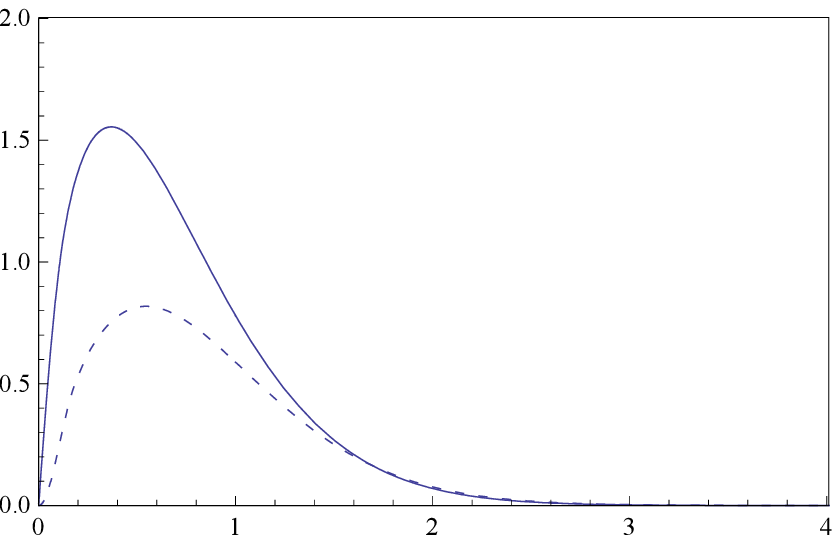}
\caption{The normalized first excited $2p\;\frac{3}{2}$ wave functions $\psi_a(x)$ 
and -$\psi_b(x)$ when $V_s(x)$ and $V_v(x)$ have equal asymptotic slopes.
Compare this solution for the massless quark case ($u$ and $d$ quarks) shown 
in Fig. \ref{fig:finalVs2p3by2}.
Compare also with the $2p\;\frac{3}{2}$ hydrogen-atom wave functions, $\psi_{a,b}(x)$ 
in Figs.\ \ref{fig:H2p32uppers} and \ref{fig:H2p32lowers} (dashed curves).}
\label{fig:VsVv2p3by2_ud}\pagebreak
\end{figure}

\begin{table}%[t]  %Table I
\begin{tabular}{|c|c|c|c|c|r|} \hline
$Z$ &\ $Z^2$\ Ry & $B$ (numerical) & $B$ (analytic) \\ \hline
10 &  1.360  & 1.3624 	& 1.3624 \\							%s 
100 & 136.0 & 161.6 & 161.6 \\
136 &\ 251.546\ \ & 433.8 & 448.3 \\ \hline
\end{tabular}
\nopagebreak
\hspace{0.25in}
\caption{Numerical results for the $1s$ ground state binding energies $B$ 
for hydrogen-like atoms with nuclear charge Z. 
The second column displays the non-relativistic eigenenergy
and the third the result of the numerical integrations.
For comparison, the analytic eigenenergies from Eq.\ (\ref{eq:analyticEgs}) are 
given in the fourth column.  Energies here are in keV.}
\hspace{0.5in}
\end{table}

\begin{table}%[c] %Table II
%\begin{center}
\begin{tabular}{|c|c|c|c|c|} \hline
   & \multicolumn{2}{c|}{\ Point Coulomb\ } & \multicolumn{2}{c|}{\ Smeared Coulomb \ } \\ \hline
$\ \alpha_s$ &\ $E$ (GeV) &\ $a_0$ &\ $E$ (GeV) &\ $a_0$\\ \hline
\ 0.0 \ &\ 0.306  \ &\ 0.195 \ &\ 0.306  \ &\ 0.195 \ \\
\ 0.2 \ &\ 0.251  \ &\ 0.244 \ &\ 0.251  \ &\ 0.237 \ \\
\ 0.4 \ &\ 0.194  \ &\ 0.343 \ &\ 0.194  \ &\ 0.305 \ \\
\ 0.6 \ &\ 0.135  \ &\ 0.551 \ &\ 0.135  \ &\ 0.417 \ \\
\ 0.8 \ &\ 0.072  \ &\ 1.052 \ &\ 0.074  \ &\ 0.611 \ \\
\ 1.0 \ &\ 0.004  \ &\ 2.653 \ &\ 0.007  \ &\ 0.977 \ \\
\ 1.2 \ &\ -.089  \ &\ 13.46 \ &\ -.156  \ &\ 1.759 \ \\ \hline
\end{tabular}
\hspace{0.25in}
\caption{Numerical results from solving the radial Dirac equastions for both $V_s(x)$ and 
a Coulombic $V_v(x)$, showing the variation of the ground state energy as a function
of $\alpha_s$.  Also shown is how the initial slope parameter $a_0$ changes with $\alpha_s$.}
%\end{center}
\end{table}


\begin{thebibliography}{99}

\bibitem{Dirac} P.\ A.\ M.\ Dirac, Proc.\ Roy.\ Soc.\ (London), {\bf A117}, 610 (1928)
and {\bf A118}, 351 (1928).

\bibitem{Bag} C.\ E.\ DeTar, J.\ F.\ Donoghue, {\it Bag Models Of Hadrons},
Ann.\ Rev.\ Nucl.\ Part.\ Sci., 235-264 (1983).

\bibitem{AWT} A.\ W.\ Thomas, S.\ Teberge, and G.\ A.\ Miller, 
Phys.\ Rev.\ D {\bf 24}, 216 (1981); 
D.\ B.\ Leinweber, A.\ W.\ Thomas, and R.\ D.\ Young,
Phys.\ Rev.\ Lett.\ {\bf 86}, 5011 (2001).

\bibitem{GMSS} T.\ Goldman, K.\ R.\ Maltman, G.\ J.\ Stephenson, Jr.,  
and K.\ E.\ Schmidt, Nucl.\ Phys.\ {\bf A481}, 621 (1988).  This 
reference is henceforth abbreviated as GMSS.

\bibitem{AWT2} K.\ Saito and A.\ W.\ Thomas, Phys.\ Lett.\ {\bf B 327}, 9 (1994);
P.\ A.\ M.\ Guichon, K.\ Saito, E.\ N.\ Rodionov, and A.\ W.\ Thomas,
Nucl.\ Phys.\ {\b A 601}, 349 (1996).

\bibitem{CorPot} E.\ E.\  Eichten et al.,  Phys.\ Rev.\ Lett.\ {\bf 34}, 369 (1975); 
Phys.\ Rev.\ D {\bf 17}, 3090 (1978);  Phys.\ Rev.\ D {\bf 21}, 203 (1980).

\bibitem{FW} L.\ L.\ Foldy and S.\ A.\ Wouthuysen, Phys.\ Rev.\ {\bf 78}, 29 (1950).

\bibitem{Darwin} C. Darwin, Proc.\ Roy.\ Soc.\ (London), {\bf A118}, 654 (1928).

\bibitem{BetheSal} H.\ A.\ Bethe and E.\ E.\ Salpeter, 
{\it Quantum Mechanics of One- and Two-Electron Atoms}
(Springer, Berlin, 1957), Sec.\ 14.

\bibitem{AuRogers} See, e.g., C.\ K.\ Au and G.\ W.\ Rogers, 
Phys.\ Rev.\ A {\bf 22}, 1820 (1980).  It was on finding this paper
that we were reminded of the coupled first-order radial Dirac equations.

\bibitem{mu_xray} See, e.g., {\it Muonic Atoms and Molecules}, ed.\ by L.\ A.\ Schaller 
and C.\ Petitjean (Birkhauser, Basel, 1993).

\bibitem{BjD} J.\ D.\ Bjorken and S.\ D.\ Drell, {\it Relativistic Quantum Mechanics}, 
(McGraw-Hill, New York, 1964), p. 55.

\bibitem{piExchg} T.\ Goldman and R.\ R.\ Silbar, Phys.\ Rev.\ C {\bf 77}, 865203 (2008).

\bibitem{Paris} M.\ Paris, Phys. Rev. C {\bf 68}, 025201(2003).

\bibitem{JF} A.\ Soares de Castro and J.\ Franklin, Int.\ J.\ Mod.\ Phys. A
{\bf 15}, 4355 (2000)

\bibitem{Critch} For the case when $r_0 = 0$, see C.\ L.\ Critchfield,
Phys.\ Rev.\ D {\bf 12}, 923 (1975).

\bibitem{AprilTalk} This interaction energy between two different quarks
often goes by the name ``color magnetic interaction,'' which is essentially 
a hyperfine-like matrix element involving the coupling between the two quark's spins.
T.\ Goldman and R.\ R.\ Silbar, talk at the April APS Meeting, 2008, Denver CO.

\bibitem{RK} E.g., George E.\ Forsythe, Michael A.\ Malcolm, and Cleve B.\ Moler. 
{\it Computer Methods for Mathematical Computations}, 
(Englewood Cliffs, NJ: Prentice-Hall, 1977). See Chapter 6.

\bibitem{MMa} Mathematica is a software product of Wolfram Research,
{\verb+http:\\www.wolfram.com+},
and its use is described by S.\ Wolfram in {\it The Mathematica Book}, 4th ed.\ 
(Cambridge University Press, Cambridge, 1999).

\bibitem{RRSwebpage} {\tt http://t16web.lanl.gov/Silbar/}.

\bibitem{hbarc} In more usual units, $\hbar c$ = 0.19732 GeV-fm, which can be used for converting 
from 1/fm to GeV.

\bibitem{Qqbar} T.\ Goldman and R.\ R.\ Silbar, ``$Q \bar{q}$ Mesons in a 
Relativistic Model'', in preparation.

\bibitem{Walecka} B.\ D.\ Serot and J.\ D.\ Walecka, {\it Advances in Nuclear Physics}, 
ed.\ by J.\ W.\ Negele and E.\ Vogt (Plenum, New York, 1986), Vol. 16, p. 1.

\bibitem{Isgur} N.\ Isgur, Phys.\ Rev.\ D {\bf 62}. 054026 (2000); 
Phys.\ Rev.\ D {\bf 62}. 014025 (2000).

\bibitem{PGG} P.\ R.\ Page, T.\ Goldman, and J.\ N.\ Ginocchio, Phys.\ Rev.\ Lett.\ 
{\bf 86}, 204 (2001).

\bibitem{PDG} C.\ Amsler et al. (Particle Data Group), Phys.\ Lett.\ {\bf B667}, 586 (2008). 

\bibitem{Eisberg} See, e.g., R.\ Eisberg and R.\ Resnick, {\it Quantum Physics of Atoms,
Molecules, Solids, Nuclei, and Particles}, (J. Wiley \& Sons, New York, 1974), Chap. 8,
pp. 310-311.

\bibitem{Klein} O.\ Klein, Z.\ Physik, {\bf 53}, 157 (1929).

\bibitem{Dombey} A.\ Calogeracos and N.\ Dombey,
{\it History and Physics of the Klein Paradox},
Contemp. Phys. {\bf 40}, 313 (1999)

\bibitem{Evans} See, e.g., R.\ Evans, {\it The Atomic Nucleus}, 
(McGraw-Hill, New York, 1955), Chap.\ 2.

\bibitem{GOR} M.\ Gell-Mann, R.\ Oakes, and B.\ Renner,  Phys.\ Rev.\ {\bf 175}, 2195 (1968).
See also H.\ G.\ Dosch and S.\ Narison, Phys.\ Lett.\ {\bf B417}, 173 (1998) and
M.\ Jamin, Phys.\ Lett.\ {\bf B538}, 71 (2002).

\bibitem{AbStAiry} M.\ Abramovitz and I.\ A.\ Stegun, {\it Handbook of Mathematical Functions},
(Dover, New York, 1965), p.\ 446.

\end{thebibliography}
\end{document}